\documentclass{gGAF2e}

\usepackage{amssymb,amsmath}
\usepackage{mathrsfs}  
\usepackage{amsfonts,amsbsy}
\usepackage{graphicx}
\usepackage{color}
\usepackage{bm}

\newcommand{\Jc}{\mathscr{J}}

\newcommand{\Lc}{\mathcal{L}}

\newcommand{\Uc}{\mathcal{U}}

\newcommand{\bv}{\boldsymbol{b}}
\newcommand{\fv}{\boldsymbol{f}}

\newcommand{\jv}{\boldsymbol{j}}

\newcommand{\uv}{\boldsymbol{u}}
\newcommand{\vv}{\boldsymbol{v}}
\newcommand{\wv}{\boldsymbol{w}}

\newcommand{\Bv}{\boldsymbol{B}}
\newcommand{\Mv}{\boldsymbol{M}}
\newcommand{\Rey}{\mathrm{Re}}
\newcommand{\Rm}{\mathrm{Rm}}

\newcommand{\cc}{\mathrm{c.c.}}
\newcommand{\half}{\tfrac{1}{2}}

\newcommand{\Bt}{\tilde{B}}

\DeclareMathOperator{\ReRe}{\mathrm{Re}}
\DeclareMathOperator{\ImIm}{\mathrm{Im}}

\newcommand\hidedetails[1]{}

\begin{document}

\title{Zonostrophic instabilities in magnetohydrodynamic Kolmogorov flow }

\author{Azza M.\ Algatheem$^\ast$\thanks{$^\ast$Corresponding author. Email:  aa965@exeter.ac.uk \vspace{6pt}}, 
Andrew D. Gilbert and Andrew S. Hillier\\\vspace{6pt}
Department of Mathematics and Statistics, University of Exeter, Exeter EX4 4QF, UK
\\\vspace{6pt}\received{\today} }

\maketitle

\begin{abstract}

\noindent
This paper concerns the classic problem of stability of the Kolmogorov flow $\uv = (0, \sin x)$ in the infinite $(x,y)$-plane. A mean magnetic field of strength $B_0$ is introduced and the MHD linear stability problem is studied for modes with a wave-number $k$ in the $y$-direction, and sometimes a Bloch wavenumber $\ell$ in the $x$-direction. The key parameters governing the problem are the Reynolds number $\nu^{-1}$, magnetic Prandtl number $P$, and dimensionless magnetic field strength $B_0$ corresponding to an inverse magnetic Mach number. The mean magnetic field can also be taken to have an arbitrary direction in the $(x,y)$-plane and a mean $x$-directed flow $U_0$ can be incorporated, in the most general formulation. 

\noindent
First the paper considers Kolmogorov flow with a $y$-directed mean magnetic field, which for convenience we refer to as `vertical'. Taking $\ell=0$, the suppression of the classic hydrodynamic instability is observed with increasing field strength $B_0$. 
A branch of strong-field instabilities occurs when the magnetic Prandtl number $P$ is less than unity, as found recently by A.E. Fraser, I.G. Cresser and P. Garaud (\emph{J. Fluid Mech.} {\bf 949}, A43, 2022). Analytical results based on eigenvalue perturbation theory in the limit $k\to0$ support the numerics for both weak-field and strong-field instabilities, and show their origin in the coupling of large-scale weakly damped modes with $x$-wavenumber $n=0$, to smaller-scale modes having $n= \pm1$. 

\noindent
The paper then considers the case of `horizontal' or $x$-directed mean magnetic field. Here the unperturbed state consists of steady, wavey magnetic field lines distorted by the underlying Kolmogorov flow, and with the driving body force balancing both viscosity and the Lorentz force. As the magnetic field is increased from zero, the purely hydrodynamic instability is suppressed again, but for stronger fields a new branch of instabilities appears. Allowing a non-zero Bloch wavenumber $\ell$ allows further instability, and  in some circumstances when the system is hydrodynamically stable, arbitrarily weak magnetic fields can give growing modes, via the instability taking place on large scales in $x$ and $y$.  Numerical results are presented together with eigenvalue perturbation theory in the limits $k, \ell\to0$. The theory gives analytical approximations for growth rates and thresholds that are in good agreement with those computed.

\end{abstract}


\section{Introduction} 

The Kolmogorov flow, a periodic flow forced at a single wavenumber, is a classic flow to study due to its simplicity and its wide applications to the study of different geophysical and astrophysical systems. Its stability to linear disturbances is a classic problem first posed by Kolmogorov and studied by \cite{meshalkin1961investigation}. These authors made use of continued fraction expansions to establish properties of the growth rate $p(k)$, where $k$ is a wavenumber in the $y$-direction, and determined a critical Reynolds number of $\Rey_c = \sqrt{2}$. Close to onset of instability, for $\Rey$ slightly larger than $\sqrt{2}$, it is large scale modes in $y$ that are destabilised; more precisely, for $\Rey = \sqrt{2} (1+3 k^2 + \cdots) $ the most unstable mode has wave number $k\ll1$. This property allows the development of amplitude equations governing the flow on large space and time scales, obtained by \cite{nepomniashchii1976stability} and \cite{sivashinsky1985weak}.  Numerical simulations by \cite{she1987metastability} showed evolution from the most unstable scale to larger scales via an inverse cascade of vortex pairings, for a large scale allowed only in the $y$-direction. For large scales in both $x$- and $y$-directions, \cite{sivashinsky1985weak} showed evolution to a large-scale flow with chaotic temporal fluctuations, further explored in \cite{lucas2014spatiotemporal,lucas2015recurrent}.

The stability problem posed by Kolmogorov is such a fundamental building block in stability theory that it has been elaborated in several studies by incorporating further physical phenomena. \cite{frisch1996large} included a $\beta$-effect, giving the gradient of a background planetary vorticity distribution; the gradient is oriented along the $x$ direction (again following our conventions rather than those of the original paper), so that it does not interact directly with the transverse, basic state Kolmogorov flow $\uv_0$. These authors derived an amplitude equation near to onset for a large scale in $y$, which they called the $\beta$--Cahn--Hilliard equation. For $\beta=0$ this reduces to the PDE of \cite{sivashinsky1985weak} and simulations show that the inverse cascade of structures to large scales in $y$ is arrested by the $\beta$-effect.
These authors characterise the fundamental instability of the Kolmogorov flow as due to a negative eddy viscosity, in other words that the large-scale $y$-dependent modes have growth rate $p  = - \nu_E k^2 + \cdots$ where the eddy viscosity $\nu_E$ changes sign from positive below the threshold $\Rey_c=\sqrt{2}$, to negative above, so destabilising the flow on large scales with the fastest growing modes determined by the next terms in this series \citep{dubrulle1991eddy}. 

In terms of  the geophysical motivation for these stability problems, any orientation of the background vorticity gradient, parameterised by $\beta$, with respect to the Kolmgorov flow is of interest. \cite{Manfroi2002stability} allow an arbitrary angle $\alpha$ between flow and gradient  in a study of linear stability and nonlinear evolution using amplitude equations generalising those of \cite{sivashinsky1985weak} and \cite{frisch1996large}. They find that the linear problem shows a delicate dependence of critical Reynolds number on angle $\alpha$ when unstable modes are allowed to adopt arbitrarily large scales in $x$ and $y$.  
Another effect of geophysical relevance that may be included is stratification. \cite{balmforth2002stratified} considered the sinusoidal flow in the $(x,z)$ plane with gravity in the $-z$ direction and the flow directed in $x$, sinusoidal in $z$. These authors determined the behaviour of  linear instabilities, depending on Reynolds, Richardson and Prandtl numbers, and derived an amplitude equation generalising that of  \cite{sivashinsky1985weak}. Simulations show that the inverse cascade of \cite{she1987metastability} is arrested by the presence of stratification over a wide range of parameters. 

Relevant to the present paper, in astrophysical applications it is natural to  introduce a magnetic field and study the coupled MHD system; as general motivation we note, for example, that the interaction between magnetic field, shear, and convection remains poorly understood in the solar tachocline  \citep{hughes2007solar}. \cite{boffetta2000large} considered the case in which a sinusoidal magnetic field (maintained by a source term in the induction equation) sits in a motionless fluid. This magnetic Kolmogorov system shows instabilities and an amplitude equation gives an inverse cascade to large scales. The recent paper \cite{fraser2022non} considers a background uniform magnetic field $\Bv_0 = (0, B_0)$ that is aligned with the Kolmogorov flow; this has no effect on the basic state flow but the elasticity of field lines affects perturbations depending on $y$, through the Lorentz force. These authors observe magnetic suppression of the instability first discussed by \cite{meshalkin1961investigation}, as one might intuitively expect, but also two new families of unstable modes which only exist in the presence of magnetic field. One family exists for magnetic Prandtl number $P<1$, for arbitrarily strong magnetic fields, provided the Reynolds number is above a threshold depending on $P$. This is studied numerically and growth rates obtained through asymptotic approximations for $k \ll1$; these authors refer to these modes as Alfv\'en Dubrulle--Frisch modes, as the instability can be linked to a change of sign of the eddy viscosity \citep{dubrulle1991eddy}.

Study of Kolmogorov flow instabilities is relevant to the formation of zonal flows in forced fluid systems, so-called `zonostrophic instability' \citep{galperin2006anisotropic}.  This process of jet formation has now been observed in many simulations, observations and experiments; see the representative studies, \cite{vallis1993generation}, \cite{read2007dynamics}, \cite{farrell2008formation}, \cite{scott2012structure}, \cite{srinivasan2012zonostrophic}, \cite{bouchet2013kinetic}, and \cite{parker2014generation}. 
Related to our work, \cite{tobias2007beta} incorporated a magnetic field aligned with the $x$-direction of a planar fluid system with a $\beta$-effect present, a vorticity gradient in $y$. The system was driven by a body force with a given characteristic spatial scale. These authors observed the formation of jets in the $x$-direction for zero magnetic field, but then 
the suppression of jets, even at quite weak field strengths $B_0$. For fixed non-dimensional $\beta$, forcing  and viscosity $\nu = 10^{-4}$, this process was explored by means of a series of runs with varying magnetic field strength $B_0$ and magnetic diffusivity $\eta$, and evidence for a threshold scaling law of $B_0^2 \sim \eta$ was observed \citep{tobias2011astrophysical}. \cite{constantinou2018magnetic} analysed Kelvin--Orr  shearing wave dynamics for Rossby/Alfv\'en waves and the interplay between Reynolds and Maxwell stresses, providing evidence for this $B_0^2 \sim \eta$ threshold for jet formation.
\cite{durston2016transport} focused on the couplings between large-scale zonal flow and zonal field in the presence of waves, calculating an effective viscosity and effective magnetic diffusivity, plus an  effective cross transport term in which current gradients can drive the zonal vorticity; this and other transport effects are discussed in \cite{chechkin1999negative}, \cite{kim2007role}, and \cite{leprovost2009turbulent}. 
\cite{parker2019magnetic} interpret the presence or otherwise of jets in terms of the competition between a positive magnetic eddy viscosity term and a negative, purely hydrodynamic eddy viscosity. Note that while these studies of zonostrophic instability have many qualitative features in common with the topic of Kolmogorov flow instability, there are key difference that make any direct comparison difficult, even of scaling laws. The reason is that the studies referred to in this paragraph use a forcing which has a given spatial scale but is random in time, and the forcing is kept fixed while other parameters such as the viscosity, magnetic diffusivity, magnetic field or $\beta$ are varied. In non-dimensional terms, the input parameter is a Grashof number (formed from forcing strength, length scale and viscosity) in these systems \citep{durston2016transport}; the Reynolds number is then a diagnostic parameter, and can vary considerably in different regimes depending on the dominant balances in the Navier--Stokes equation between  the forcing term, inertial term, viscous term  and Lorentz force. However for stability of Kolmogorov flow, the basic state is fixed while the forcing is adjusted to maintain this: the input parameter is a Reynolds number, instead.

In the present paper we return to the classic set-up of steady, planar, Kolmogorov flow  $\uv_0 = (0, \sin x)$ and consider the effect on its stability from magnetic field in the $x$- and $y$-directions. We find it convenient to refer to magnetic field in the $y$-direction, parallel to the flow as `vertical field' and magnetic field in the $x$-direction, aligned with possible jet formation, as `horizontal field' (even though gravity/stratification are not  involved in our study). In \S 2 we set up the equations solved for linear perturbations with vertical magnetic field and in section 3 present numerical and analytical results, showing growth rates, thresholds and unstable mode structure. This section has common elements with the recent paper  \cite{fraser2022non} (published while the present paper was in preparation); however we find it useful to set out the numerical results to compare with the later horizontal field case, and we present new analytical results in \S 3.1 for the `weak vertical field branch'. The `strong vertical field branch' in \S 3.2 is a primary focus for \cite{fraser2022non}, and we give an alternative, matrix-based derivation of the asymptotic growth rate they obtain. 
Section 4 sets up the equations for horizontal magnetic field, with numerical results supported by analytical approximations in the limit $k\to0$ given in \S5, together with the case of non-zero Bloch wavenumber $\ell$, and \S6  offers concluding discussion. Further analytical and numerical results will appear in \cite{mephd2023}. To keep the main body of the paper compact, we have developed analytical theory in appendices, building up in order of complexity rather than in the order in which the results are used. The method employed is perturbation theory for eigenvalues and eigenvectors of a matrix; naturally this is equivalent to methods used by other authors, but we find that is a systematic way of handling problems of increasing complexity, and gives insight into how  couplings between individual flow and field modes can drive an instability.

%
%

%
%
 

\section{Governing equations: vertical field}

Our starting point is the system of equations for incompressible, constant density MHD, written in the form 
\begin{align}
& \partial_t \uv   + \uv\cdot\nabla\uv = - \nabla p + \jv \times \bv  +  \nu \nabla^2 \uv + \fv, \label{eqMHDu} \\
& \partial_t  \bv  + \uv \cdot \nabla \bv  = \bv  \cdot \nabla\uv + \eta \nabla^2 \bv , \label{eqMHDb} \\
&  \nabla \cdot \uv = 0,  \quad \nabla \cdot\bv = 0 , \quad \jv = \nabla \times \bv. \label{eqMHDsol}
\end{align}
Here the magnetic field is represented in velocity units, and $\fv$ is an externally imposed rotational body force, used to maintain a given basic state for the system. In dimensional units this basic state is the Kolmogorov flow, namely the unidirectional flow in the $(x,y)$-plane specified by 
\begin{equation}
\uv_0= \Uc (0, \sin (x/\Lc) ) .  
\end{equation}
We use the length $\Lc$ and velocity $\Uc$ as the basis of our non-dimensionalisation, which results in the same equations (\ref{eqMHDu}--\ref{eqMHDsol}) above but having $\nu$ now identified as an inverse Reynolds number, with $\Rey = \nu^{-1}$, and $\eta$ as an inverse magnetic Reynolds number, with $\Rm = \eta^{-1}$. The  non-dimensional Kolmogorov flow  is then 
\begin{equation}
\uv_0 =  (0, \sin x ) . 
\label{eqbasicnondim}
\end{equation}
Note that we drop the $z$-components of vectors where we can. A key quantity we will need is the magnetic Prandtl number 
\begin{equation}
P = \Rm /  \Rey = \nu/\eta. 
\end{equation}

We will consider flows $\uv$ and magnetic fields $\bv$ lying in the $(x,y)$-plane, independent of $z$. For this we use a stream function $\psi$ and vector potential $a$ defined by 
\begin{equation}
\uv = (\psi_y, -\psi_x), \quad 
\bv = (a_y, - a_x).
\end{equation}
The governing equations may then be writtten in terms of the evolution of a scalar vorticity $\omega$, and $a$:
\begin{align}
& \partial_t \omega + \! \Jc (\omega,\psi) =  \! \Jc (j, a) + \nu  \nabla^2 \omega  +  g, \label{eqMHDomega} \\
& \partial_t  a   + \!\Jc (a,\psi) = \eta  \nabla^2 a , \label{eqMHDa} \\
&  \omega = - \nabla^2 \psi, \quad j = - \nabla^2 a  . \label{eqMHDrel}
\end{align}
Here $\Jc$ is the Jacobian of two functions  in the plane and $g$ is the $z$-component of the curl of the body force $\fv$.

\begin{figure}
\centering
\hspace{-.5cm}  
\includegraphics[scale=0.5]{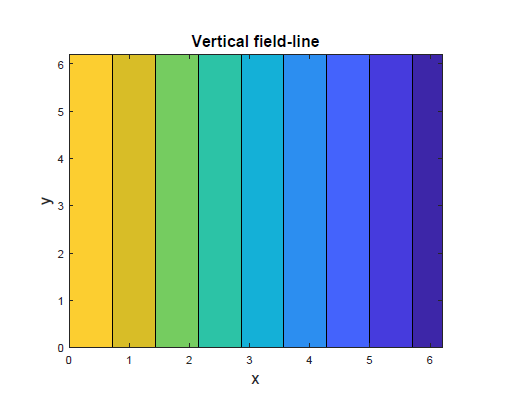}
\includegraphics[scale=0.5]{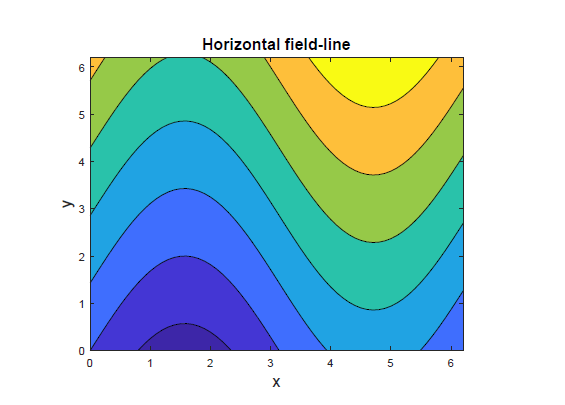}
\caption{The magnetic field basic state for (a) vertical field (in \S\S2, 3), (b) horizontal field (in \S\S4, 5), with $B_0=0.7$, $\eta = 0.5$. In each case fields lines are depicted as contours of the corresponding vector potential $a_0$, with $\bv_0 = ( \partial_y a_0, - \partial_x a_0)$. }
 \label{fig-basic-state}
\end{figure}

We begin with the study of the stability of  Kolmogorov flow in the presence of vertical magnetic field (that is, $y$-directed field) of strength $B_0$ \citep{fraser2022non}. Aiming for the most general set-up we  also include  a uniform horizontal flow (that is, an $x$-directed flow) of strength $U_0$. We therefore adopt the following steady solution of the equations as our basic state,
\begin{equation}
\uv_0 =  (U_0, \sin x),   \quad \bv_0 = (0, B_0), \quad \fv = (0, \nu \sin x + U_0 \cos x ),
\end{equation}
or in our scalar-based formulation
\begin{equation}
 \psi_0 =U_0 y + \cos x, \quad
  \omega_0 = \cos x, \quad
a_0 = - B_0 x  , \quad
j_0 = 0 , \quad g = \nu \cos x - U_0 \sin x  .
\end{equation}
The basic state magnetic field is shown in figure \ref{fig-basic-state}(a). The stability problem is parameterised by the four quantities $\{ \nu, B_0, P, U_0\}$. Note that while the mean horizontal flow specified by the parameter $U_0$ could be removed by a Galilean transformation, the Kolmogorov flow would then become a travelling wave. Thus, given we always take a steady Kolmogorov flow in the form (\ref{eqbasicnondim}), the effect of $U_0$ cannot be eliminated by this means.

To study  the stability of this basic state we linearise, replacing
\begin{equation}
 \psi = \psi_0 + \psi_1 + \cdots, \quad \omega = \omega_0 + \omega_1 + \cdots, \quad 
 a = a_0 + a_1 + \cdots , \quad j = j_0 + j_1 + \cdots, 
\end{equation}
and, droppping the subscript $1$, we deduce the linear system
\begin{align}
& \partial_t \omega + U_0 \,\omega_x + \sin x \, (\omega_y-  \psi_y)  = B_0\, j_y  + \nu  \nabla^2 \omega  , \label{eqMHDomegalin1} \\
& \partial_t  a   + U_0\, a_x  + \sin x \, a_y = B_0\, \psi_y + \eta  \nabla^2 a , \label{eqMHDalin1} \\
&  \omega = - \nabla^2 \psi, \quad j = - \nabla^2 a  . \label{eqMHDrellin1}
\end{align}

We now expand the fields in Fourier modes in $x$ as 
\begin{align}
(\psi, \omega, a, j)  & = e^{pt + i \ell x + iky} \, \, \sum_{n} \, \, \, ( F_n, G_n, H_n, J_n) \, e^{inx}  + \cc  
\label{eqfourier} 
\end{align}
Here $p$ is the complex growth rate of the mode, $k$ is the wavenumber in the $y$-direction with $k>0$ without loss of generality, and $\ell$ is a Bloch or Floquet wavenumber in the $x$-direction satisfying $-1/2 < \ell \leq 1/2$. 

Substituting these series into the linear equations (\ref{eqMHDomegalin1}--\ref{eqMHDrellin1}) results in an infinite system of equations. For $\ell=0$ these may be written in the form:
\begin{align}
pG_n  = - [ \nu(n^2+k^2) +  in U_0] G_n 
& +\frac{k}{2} \left(\frac{1}{(n-1)^2+k^2}- 1\right)G_{n-1} 
\label{eqGnvert}\\
& -\frac{k}{2} \left(\frac{1}{(n+1)^2+k^2}-1 \right)G_{n+1} 
+ik B_0(n^2 + k^2)H_n, 
\notag\\
pH_n =- [ \eta(n^2 + k^2) + in  U_0] H_n   &  -\frac{k}{2}\, H_{n-1}+\frac{k}{2}\, H_{n+1}+\frac{ikB_0}{n^2 + k^2}\, G_n , 
\end{align}
and for $\ell\neq 0$ we simply replace $n \to n + \ell$ wherever it appears (except as a subscript). This provides an eigenvalue problem giving a discrete set of eigenvalues with a dependence $p(k, \ell, \nu, B_0, P, U_0)$ in general, and the real part of the growth rate is unchanged on the replacement $(k, \ell) \to (-k , -\ell)$. 

For a numerical solution we restrict $-N \leq n \leq N$ for some integer $N$ (typical values are $16$, $32$) and solve a discrete matrix problem written in the  pentadiagonal form 
\begin{equation}
p \begin{pmatrix}  
\vdots \\[4pt] G_n \\[6pt]  H_n \\[4pt] \vdots 
\end{pmatrix} =
\left(
\begin{array}{ccc|cc|ccc}
\ddots & \ddots & \ddots &\ddots &\ddots &\ddots &\ddots &\ddots    \\ \hline
\ddots &   \otimes & 0 & \otimes & \otimes & \otimes & 0  & \ddots\\
\ddots &  0 & \otimes &   \otimes & \otimes  &  0 &  \otimes  & \ddots \\  \hline
  \ddots & \ddots &\ddots &\ddots &\ddots &\ddots &\ddots &\ddots   \\
\end{array}
\right)
 \begin{pmatrix} 
 \vdots \\[4pt] G_n \\[6pt]  H_n \\[4pt] \vdots 
 \end{pmatrix}, 
 \label{eqmatrixsystem}
\end{equation}
where $\otimes$ denotes the only non-zero entries. At a specified truncation $N$ the $(4N+2)\times (4N+2)$ matrix is set up in Matlab, and an eigenvalue $p$ with maximum real part is calculated. For a given parameter set $\{ \nu, B_0, P, U_0\}$ the maximum real growth rate is defined as 
\begin{equation}
\ReRe p_{\max} ( \nu, B_0, P, U_0) = \max_{k,\ell} \, \ReRe p(k, \ell, \nu, B_0, P, U_0), 
\end{equation}
with the maximum taken over a grid of $k$ and $\ell$ values. 

In what follows we will start by taking $U_0=0$, $\ell=0$ and only vary the vertical wavenumber $k$. The maximisation is then taken over a finite range of $k$-values, typically 100 values in the range $0\leq k \leq 1$, and any complex eigenvalues appear in  complex conjugate pairs. We let $k_{\max}(\nu, B_0, P)$ be the corresponding maximising wave number, and we attach the appropriate (zero or positive) imaginary part to give $p_{\max}(\nu, B_0, P)$ as the (maximum) complex instability growth rate. It is then often instructive to plot $\ReRe p_{\max}$, $\ImIm p_{\max}$ and $k_{\max}$. 


\section{Numerical and analytical results: vertical field}


\begin{figure}
\centering
\hspace{-.5cm}  
(a)\includegraphics[scale=0.45]{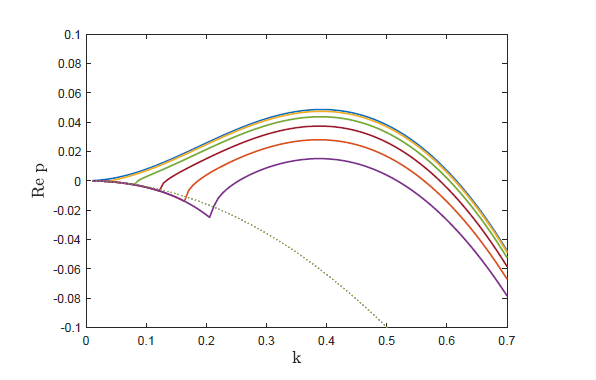}
(b)\includegraphics[scale=0.45]{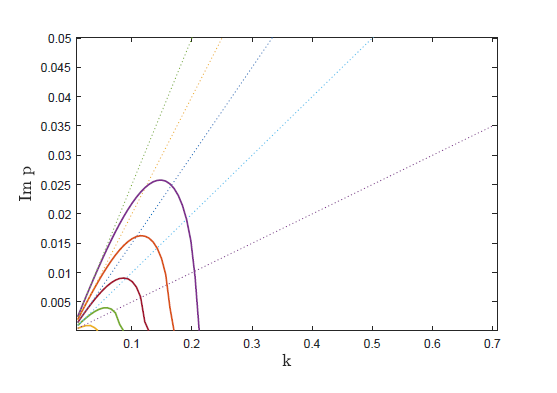}
\caption{Instability growth rate $p$ for vertical magnetic field as a function of wave number $k$ (with $U_0$, $\ell=0$) for 
$\nu = \eta =0.4$ ($P=1$), and $B_0 =0$ (blue), $B_0=0.05 $ (orange), $B_0=0.10 $ (green), $B_0=0.15$ (dark red), $B_0=0.20$ (red) and $B_0=0.25 $ (purple). 
Panels (a) and (b) show $\ReRe p$ and $\ImIm p$ respectively, and dotted curves show the Alfv\'en wave branch in (\ref{eqalfvenwaves}). 
}
 \label{fig-vertical-p}
\end{figure}

We use the numerical code as described above to produce eigenvalues so that we can explore the dependence on parameters. Our starting point is to  investigate the effect of increasing the vertical magnetic field strength $B_0$ on the classic hydrodynamic instability of Kolmogorov flow. 

\subsection{Weak vertical field branch} 

Figure \ref{fig-vertical-p}(a) shows the real part of the growth rate $p(k, \nu, B_0,P)$ for $\nu = \eta = 0.4$ and so $P=1$, plotted against $k$ for given values of the magnetic field strength $B_0$. 
Here $B_0$ is increased from zero in steps of $0.05$ as we read down the family of curves. The top curve relates to the purely hydrodynamic case. As we increase $B_0$ we note two effects: first the peak is reduced, in other words the magnetic field acts to suppress the instability, as found by \cite{fraser2022non}. Secondly, for large scales, namely small $k$, a new branch of decaying modes appears, with growth rates largely independent of field strength. Figure \ref{fig-vertical-p}(b) shows the imaginary part of the growth rate $p$. This is zero for the purely hydrodynamic case and remains zero for this branch as it is suppressed by the field. The new branch for low $k$ has a non-zero imaginary part which becomes more prominent as $B_0$ is increased and we read up the curves. Some investigation shows that the new branch is in fact a damped Alfv\'en wave on the vertical magnetic field. For zero background flow $\uv_0$, a vertical field supports Alfv\'en waves with 
\begin{equation}
p =  \pm ik \sqrt{B_0^2 - \tfrac{1}{4} (\nu-\eta)^2 k^2} - \tfrac{1}{2} (\nu + \eta) k^2, 
\label{eqalfvenwaves}
\end{equation}
and the real and imaginary parts are shown dotted in figure \ref{fig-vertical-p}. Since the Alfv\'en waves are modified by the background Kolmogorov flow, the agreement is not exact, but this is clearly the origin of these small-$k$ damped modes. 

\begin{figure}
\centering
\hspace{-.5cm}  
(a)\includegraphics[scale=0.4]{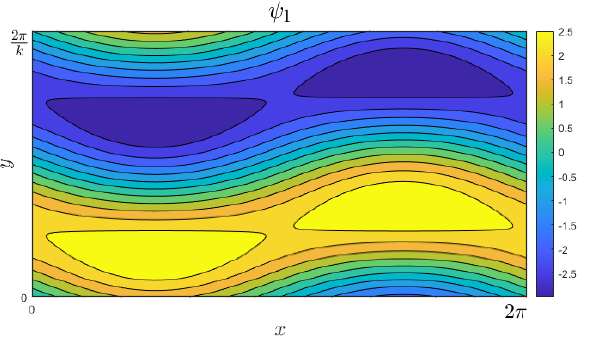}
(b)\includegraphics[scale=0.4]{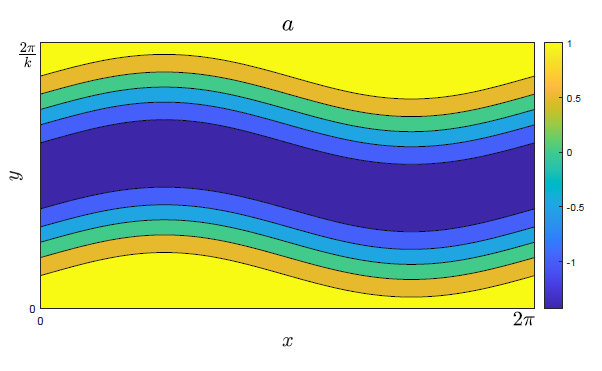}
 \caption{Structure of a typical unstable mode, with $B_0=0.25$, $\nu=\eta = 0.4$, $k=0.4$; (a) shows the stream function $\psi$ and (b) the vector potential $a$. }
 \label{fig-vertical-mode}
\end{figure}

A typical unstable mode  is shown in figure \ref{fig-vertical-mode} for parameter values corresponding to the peak in the lowest curve in figure \ref{fig-vertical-p}(a), that is for the strongest field used. We observe the perturbation streamfunction $\psi$ in panel (a) showing clear zonostrophic jets, and corresponding changes to the magnetic vector potential in panel (b); \cite{fraser2022non} refer to these as sinuous Kelvin--Helmholtz modes. As the instabilities we observe here are obtained from the hydrodynamic problem as we increase $B_0$ from small values, we refer to this as the \emph{weak vertical field branch}, to be contrasted with a strong field branch we encounter shortly. 


\begin{figure}
\centering
\hspace{-.5cm}  
(a)\includegraphics[scale=0.5]{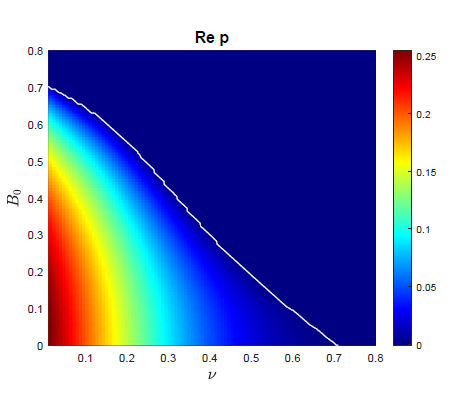}
(b)\includegraphics[scale=0.5]{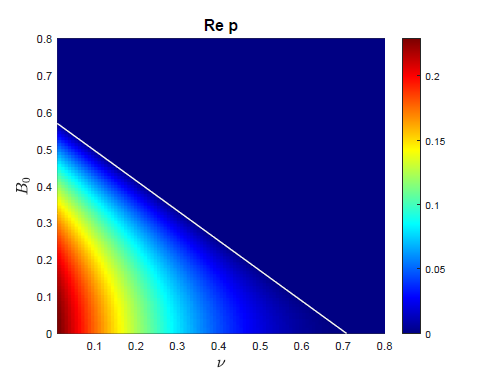}
\caption{Instability growth rate $\ReRe p_{\max}$ for vertical field plotted in the $(\nu, B_0)$ plane for $P=1$, $\ell=0$, 
$U_0=0$. Panel (a) shows the numerical computation of growth rates with the threshold $\ReRe p_{\max} = 0 $ given by a white curve, and 
panel (b) the analytical maximum growth rate from (\ref{eqappCpmax}) and threshold from (\ref{eqappCpthreshold}).}
\label{fig-vertical-nu-B0}
\end{figure}

Having seen a particular example of how the magnetic field suppresses the hydrodynamic instability by plotting $p(k,\nu, B_0,P)$, we now show results where we maximise over $k$ for each set of the parameters. Figure \ref{fig-vertical-nu-B0}(a) shows numerical results for $\ReRe p_{\max}( \nu, B_0, P)$ with $P=1$ as a colour plot across the $(\nu,B_0)$-plane. The white curve shows the threshold for instability (actually set for $\ReRe p_{\max}$ having a small positive value). The horizontal axis $B_0=0$ is the hydrodynamic case, where the white curve crosses at $\nu_c = 1/\sqrt{2}$. Instability occurs in the region below the white curve, and we can see that it is suppressed as $B_0$ increases, up to the point where $B_0\simeq 0.7$ and the instability is entirely eliminated. We do not show $\ImIm p_{\max} $, which is zero within the region of instability. 

We can develop perturbation theory \cite[as in, for example,][]{frisch1996large,Manfroi2002stability} to calculate approximate growth rates valid for $k\to0$. In appendix C we give the details. One key result is the formula (\ref{eqappCp}), 
\begin{equation}
p (k, \nu, B_0, P) =  -  \frac{\sqrt{2}\, P}{1+P^2} \,   B_0^2  +2 \biggl(\frac{1}{\sqrt{2}} - \nu\biggr) k^2 - \frac{3}{\sqrt{2}}\,  k^4  + \cdots , 
\end{equation}
giving the growth rate, and showing clearly how the effect of the magnetic field is to suppress the hydrodynamic $B_0=0$ instability  (as seen in figure \ref{fig-vertical-p}(a)). For unstable modes it is necessary that $\nu < 1 /\sqrt{2}$ and in this case maximising the growth rate over values of $k$ gives
\begin{equation}
p_{\max} (\nu , B_0, P) = -  \frac{\sqrt{2}\, P}{1+P^2} B_0^2 + \frac{\sqrt{2}}{3} \biggl(\frac{1}{\sqrt{2}} - \nu\biggr)^2 , \quad k_{\max}^2  =  \frac{\sqrt{2}}{3} \biggl(\frac{1}{\sqrt{2}} - \nu\biggr). 
\label{eqappCpmax}
\end{equation}
Putting $p_{\max}=0$ gives the threshold of instability as the straight line in the $(\nu, B_0)$-plane:
\begin{equation}
B_0 (\nu, P) = \sqrt{\frac{1+P^2}{3P}}\, \biggl(\frac{1}{\sqrt{2}} - \nu\biggr) .
\label{eqappCpthreshold}
\end{equation}
Figure \ref{fig-vertical-nu-B0}(b) shows the theoretical growth rate and the threshold marked by a white (straight) line, showing good agreement with the full numerics. The perturbation theory is developed about the point $\nu_c = 1/\sqrt{2}$, $B_0=0$ of the onset of the hydrodynamic instability. Hence the agreement is particularly good near this point; elsewhere the theory gives results that are qualitatively correct only, for example the theoretical value of $B_0$ which suppresses instability for all $\nu$ is given by 
\begin{equation}
B_0 \simeq B_* = \sqrt{ \frac{1+P^2}{6P}} \,  , 
\label{eqvertweakfieldbstar}
\end{equation}
with $B_*\simeq 0.58 $ for $P=1$ in panel (b), which is a little lower than the actual value $B_* \simeq 0.7$ seen for the numerical results in panel (a).



\begin{figure}
\centering
\hspace{-.5cm}  
\includegraphics[scale=0.5]{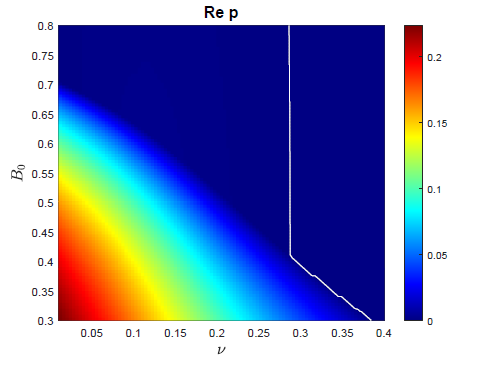}
\includegraphics[scale=0.5]{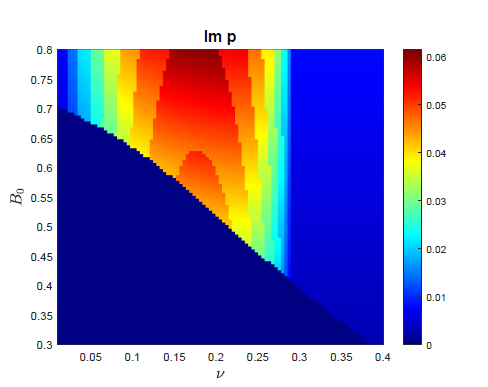}
 \caption{Shown are (a) instability growth rate $\ReRe p_{\max}$ and (b) frequency $\ImIm p_{\max}$ for vertical field plotted in the $(\nu, B_0)$ plane, with $P=\tfrac{1}{2}$. The results shown are obtained computationally and the white curve in panel (a) shows zero growth rate.  }
 \label{fig-strong-vertical-nu-B0}
\end{figure}

\subsection{Strong vertical field branch} 

Although the magnetic field acts to suppress the instability for magnetic Prandtl number $P=1$, this is not the whole picture, and investigations for $P<1$ show the presence of a \emph{strong vertical field branch}, as found by \cite{fraser2022non}. Figure \ref{fig-strong-vertical-nu-B0} shows (a) the real part and (b) imaginary part of the growth $p_{\max} (\nu, B_0, P)$ for $P=1/2$, that is $\eta = 2 \nu$. The threshold $\ReRe p_{\max} = 0 $ is shown as a white curve in both figures. Looking from the bottom of the panel \ref{fig-strong-vertical-nu-B0}(a) (increasing $B_0$) we see that the curving white line, showing the weak field branch in panel \ref{fig-vertical-nu-B0}(a), turns to become a near-vertical line, demarcating a new branch with non-zero frequency $\ImIm p_{\max}$ evident in panel \ref{fig-strong-vertical-nu-B0}(b).

This strong vertical field branch is analysed in appendix B, using a scaling in which $B_0 = O(k^{-1})$ as $k\to0$. The pertubation theory then involves a leading order undamped Alfv\'en wave with frequency $p_0 = ikB_0 = O(1)$.  The coupling of this wave with the flow field leads to potential instability with a growth rate  given in (\ref{eqstrongvertfieldRep}) as: 
\begin{equation}
\ReRe p =   \frac{\tfrac{1}{4}   \nu \eta (\eta - \nu )k^2 }{\nu^2 \eta^2 + k^2 B_0^2  (\nu + \eta)^2 } -  \tfrac{1}{2} (\nu + \eta) k^2 + \cdots .
\label{eqstrongvertfieldRep0} 
\end{equation}
An equivalent expression is found in \cite{fraser2022non} by approximating a quartic dispersion relation. Evidently we need $\eta>\nu$ for instability, in other words $P<1$; the instability of the large-scale Alfv\'en wave appears to take a double-diffusive form. As discussed in the appendix, the stability threshold (\ref{eqstrongvertfieldRep}) of \cite{fraser2022non}  is also obtained from this equation as 
\begin{equation}
\nu < \nu_* = \sqrt{ \frac{P}{2} \, \frac{1-P}{1+P} } \, .
\label{eqstrongvertfieldRep1} 
\end{equation}
For example if $P=1/2$ then $ \nu_* =  1/ \sqrt{12} \simeq 0.28$, in good agreement with the vertical white lines in figure \ref{fig-strong-vertical-nu-B0}. Thus the instability persists for arbitrarily large magnetic fields, provided the viscosity is below this Prandtl-number dependent threshold, in other words provided the Reynolds number $\Rey = \nu^{-1} > \nu_*^{-1}$.
 
%


\begin{figure}
\centering
\hspace{-.5cm}  
(a)\includegraphics[scale=0.55]{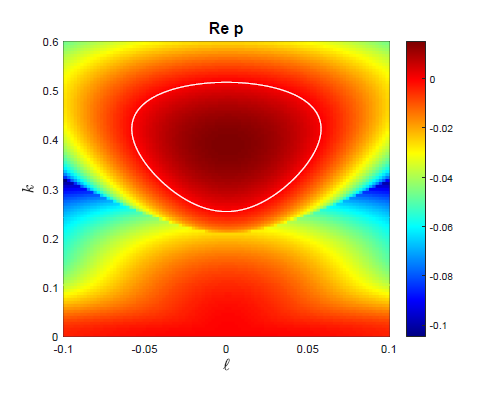}
(b)\includegraphics[scale=0.53]{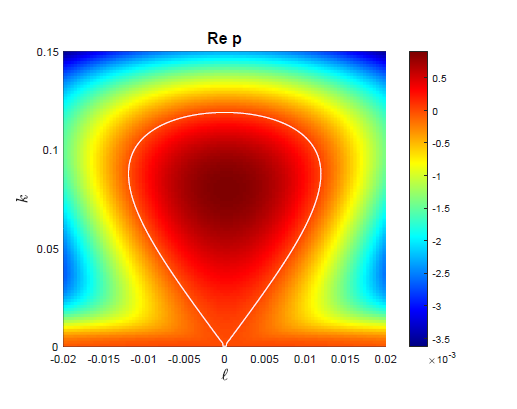}
 \caption{Instability growth rate $p$ for vertical field as a function of $(\ell,k)$ for (a)  $\nu =\eta=0.4$ ($P=1$) with $B_0 = 0.25$, and  (b) $\nu =0.2$, $\eta=0.4$ ($P=0.5$) with $B_0 = 0.7$. }

 \label{fig-vertical-field-jewel}
\end{figure}

All the above results have been taken for Bloch wave number $\ell=0$. Introducing $\ell$ brings in an extra degree of freedom and allows the possibility of new instabilities. However in the vertical field case increasing $|\ell|$ from zero has only a stabilising effect \citep{fraser2022non}. For example figure \ref{fig-vertical-field-jewel} shows growth rates in the $(\ell, k)$-plane for weak and strong field cases in panels (a) and (b). The vertical axis gives $\ell=0$ and we see an island of unstable modes in each case.  However to either side of the vertical axis, the growth rates are diminished and allowing $\ell\neq0$ has little impact. For this reason we will not consider $\ell$ further for the vertical field case, except to mention that the theory in appendix B may be extended to incorporate $\ell \neq0$, as detailed in \cite{mephd2023}.


\section{Governing equations: horizontal field, with $U_0=0$}

In this section we study the stability of  Kolmogorov flow in the presence of  horizontal field $B_0$, and to reduce the complexity of the problem we restrict to the case of zero horizontal mean flow $U_0=0$. We thus adopt the following basic state, a steady solution of the equations (\ref{eqMHDu}--\ref{eqMHDsol}): 
\begin{equation}
\uv_0 =  (0, \sin x),   \quad \bv_0 = (B_0 , \eta^{-1} B_0 \cos x), \quad \fv = (0, (\nu+ \eta^{-1} B_0^2)  \sin x ) , 
\label{eqhorizbasicstate}
\end{equation}
or in the scalar formulation
\begin{equation}
 \psi_0 = \cos x, \quad
  \omega_0 = \cos x, \quad
a_0 =  B_0 y - \eta^{-1} B_0 \sin x , \quad
j_0 =  - \eta^{-1} B_0 \sin x , \quad 
g = ( \nu  + \eta^{-1} B_0^2 )  \cos x .
\end{equation}
The basic state magnetic field is shown in figure \ref{fig-basic-state}(b), with horizontal field lines distorted by the background Kolmogorov flow, becoming increasingly extended in the limit of small $\eta$.  Note, to pick up a comment in the introduction, that the  body force required to maintain the Kolmogorov flow increases with field strength $B_0$ and with magnetic Reynolds number $\Rm = \eta^{-1}$, unlike in many large-scale simulations of zonostrophic instability, where the magnitude of a random forcing is held fixed, while other parameters are varied.  

The stability problem is parameterised by $\{ \nu, B_0, P, U_0=0\}$.  The corresponding linear system is
\begin{align}
& \partial_t \omega + \sin x \, (\omega_y-  \psi_y)  = B_0\, j_x  + \eta^{-1} B_0 \cos  x \, (j_y-a_y) +  \nu  \nabla^2 \omega  , \label{eqMHDomegalin} \\
& \partial_t  a   +  \sin x \, a_y = B_0\, \psi_x + \eta^{-1} B_0 \cos x\,  \psi_y + \eta  \nabla^2 a , \label{eqMHDalin} \\
&  \omega = - \nabla^2 \psi, \quad j = - \nabla^2 a  , \label{eqMHDrellin}
\end{align}
where the fields represent the perturbation to the basic state. The resulting equations for the Fourier modes in $x$ are 
\begin{align}
pG_n  & = -\nu(n^2+k^2) G_n
 +\frac{k}{2} \left(\frac{1}{(n-1)^2+k^2}- 1\right)G_{n-1} 
 -\frac{k}{2} \left(\frac{1}{(n+1)^2+k^2}-1 \right)G_{n+1} \notag\\
&  +in B_0(n^2 + k^2)H_n 
+ \frac{ikB_0}{2\eta} \bigl[ (n-1)^2 + k^2-1\bigr] H_{n-1} 
+ \frac{ikB_0}{2\eta} \bigl[ (n+1)^2 + k^2-1\bigr] H_{n+1} , 
\label{eqGnhoriz}
\\
pH_n & =- \eta(n^2 + k^2) H_n    -\frac{k}{2}\, H_{n-1}+\frac{k}{2}\, H_{n+1} \notag \\
& +\frac{inB_0}{n^2 + k^2}\, G_n 
+ \frac{ikB_0}{2\eta}  \frac{1}{ (n-1)^2 + k^2} \, G_{n-1} 
+ \frac{ikB_0}{2\eta}  \frac{1}{ (n+1)^2 + k^2} \, G_{n+1} ,  
\label{eqHnhoriz}
\end{align}
for $\ell=0$ and, as elsewhere, for $\ell\neq0$ we replace $n$ by $n+ \ell$. This infinite system of linear equations may then be truncated and set up as a matrix eigenvalue problem, analogously to that in (\ref{eqmatrixsystem}) for vertical field; the matrix now takes a heptadiagonal form. 


\section{Numerical and analytical results: horizontal field}

We have used Matlab to obtain growth rates $p(k, \ell, \nu, B_0, P)$ (here $U_0=0$) and we focus first on the case $\ell=0$. 


\begin{figure}
\centering
\hspace{-.5cm}  
\includegraphics[scale=0.5]{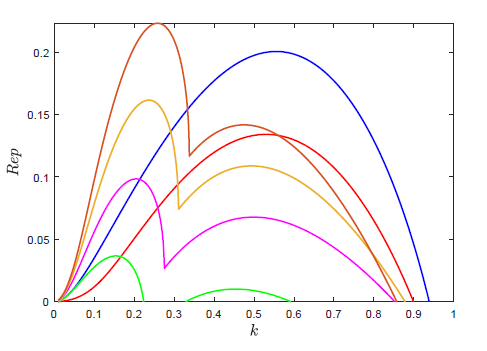}
\includegraphics[scale=0.5]{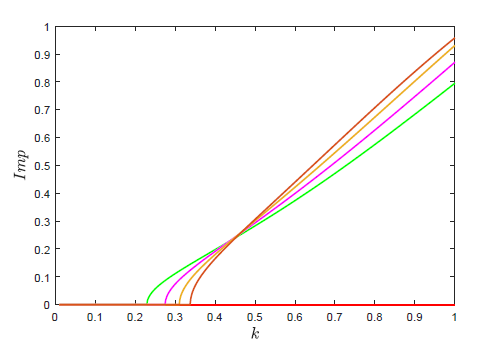}
 \caption{Instability growth rate $p$ for horizontal field as a function of $k$ (and $\ell=0$) for $\nu = \eta =0.1$ ($P=1$), with $B_0 = 0$ (blue), $0.05$ (red), $0.1$ (green), $0.15$ (purple), $0.20$ (orange) and $0.25$ (dark orange).  Panels (a) and (b) show $\ReRe p$ and $\ImIm p$ respectively.}
 \label{fig-horiz-p}
\end{figure}
 
\begin{figure}
\centering
\hspace{-.5cm}  
(a)\includegraphics[scale=0.5]{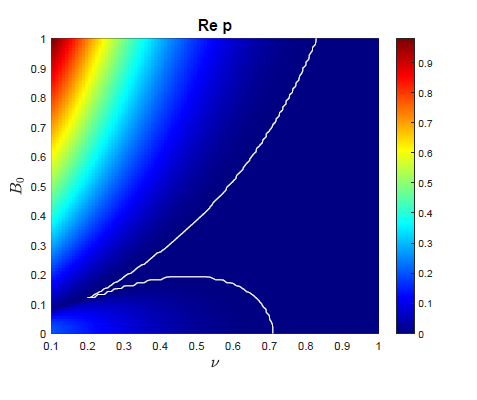}
(b)\includegraphics[scale=0.5]{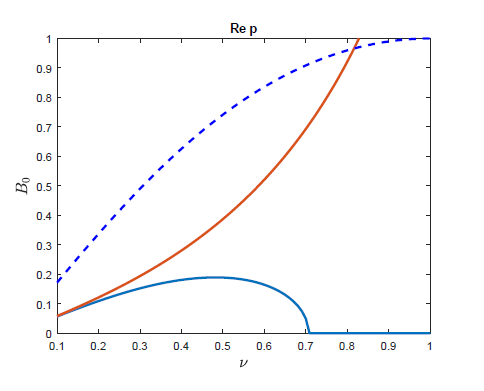}
 \caption{Instability growth rate $\ReRe p$ for horizontal field plotted in the $(\nu, B_0)$ plane for $P=1$, $\ell=0$, $U_0=0$. Panel (a) shows the numerical computation of growth rates with the threshold given by a white curve, and panel (b) the analytical thresholds from (\ref{eqG0threshold}) for the flow branch (blue), and from (\ref{eqH0threshold}) for the field branch (dark orange). In panel (b) the dashed line is the threshold (\ref{eqhorizB0ellthreshold}) for $\ell\neq0$ instabilities discussed later. }
 \label{fig-horiz-nu-B0}
\end{figure}

\subsection{Instability for Bloch wave number $\ell=0$} 

With zero Bloch wave number $\ell$, the instability has periodicity $2\pi$ in the $x$-direction and $2\pi/k$ in the $y$-direction. Figure \ref{fig-horiz-p} shows plots of the growth rate $p(k, \nu, B_0, P)$ against $k$ for $\nu = \eta = 0.1$ ($P=1$) and $B_0$ increasing as detailed in the caption. Focusing on the real part of $p$ in panel (a) we observe that the magnetic field initially suppresses the instability, going from the blue $B_0=0$ curve to the lower, red $B_0=0.05$ curve. However increasing $B_0$ further, the green $B_0 = 0.1$ curve reveals a double-peaked growth rate and then these two peaks increase as $B_0$ is increased, as indicated in the figure caption. For these stronger fields, the second peak is the lower of the two, and is associated with non-zero imaginary part $\ImIm p$ of the growth rate, as shown in panel (b), while the dominant instability of the first peak has $\ImIm p = 0 $. In fact for all our studies of instability for horizontal field with $U_0=0$ and $\ell=0$, we observe that the dominant instability is always direct or non-oscillatory, that is $\ImIm p = 0 $. 

To give a more global picture of these results for horizontal field, we now show $p_{\max}(\nu, B_0, P)$ as a colour plot in the $(\nu, B_0)$-plane with $P=1$ in figure \ref{fig-horiz-nu-B0}(a), with the white line denoting the instability threshold $\ReRe p = 0 $. For modest magnetic fields we observe the suppression of the purely hydrodynamic instability as in the case of vertical field in figure~\ref{fig-vertical-nu-B0}. However as $B_0$ is increased another branch of instability emerges from the bottom left of the panel \ref{fig-horiz-nu-B0}(a) and shows increasing growth rates, particularly for smaller viscosities $\nu$. This new branch of instabilities is perhaps not surprising \citep{durston2016transport}, given that the basic state horizontal field in (\ref{eqhorizbasicstate}) and depicted in figure \ref{fig-basic-state}(b) has a wavey structure, and for $P=1$ becomes increasingly convoluted as $\eta = \nu$ is decreased. Other values of $P$ give plots that are similar in structure. 

To gain analytical results and understanding, in appendix A we discuss perturbation theory for the horizontal field system, taking the limit $k\to0$ while retaining $B_0$ and other parameters of order unity. The resulting leading order equations involving $G_0$, $G_{\pm 1}$, $H_0$ and $H_{\pm1}$ split into the two independent $3\times 3$ matrix systems giving the two branches of instability evident in figure \ref{fig-horiz-nu-B0}(a). We discuss them in turn.

\begin{figure}
\centering
\hspace{-.5cm}  
\includegraphics[scale=0.4]{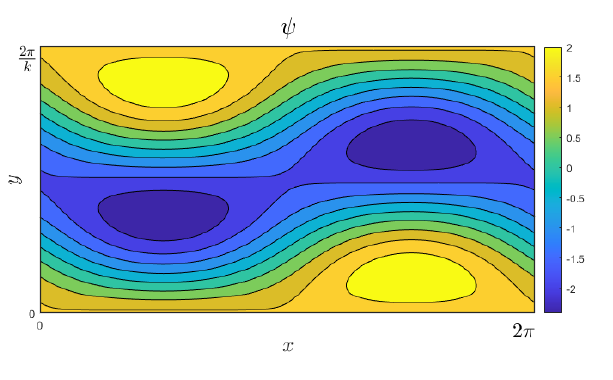}
\includegraphics[scale=0.4]{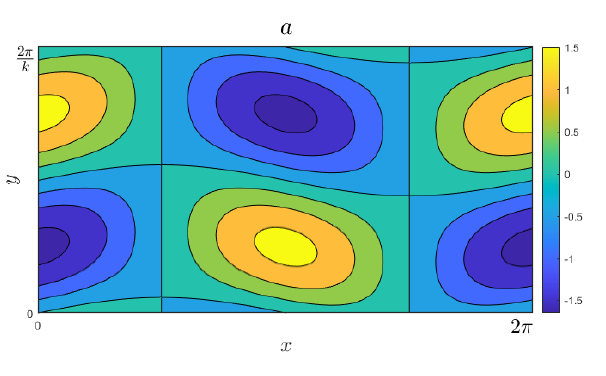}
 \caption{A typical unstable mode, with $B_0=0.05$, $\nu=\eta = 0.1$, $k=0.5$, from the flow or $G_0$ branch; (a) shows the stream function $\psi$ and (b) the vector potential $a$.}

 \label{fig-horiz-G0-mode}
\end{figure}

The first system involves $G_0$ and not $H_0$, in other words is dominated by a large-scale flow and not a large-scale field. We call this the $G_0$ or \emph{flow branch} of the horizontal field instability. Analysis gives equation (\ref{eqphorizG0}), which we reproduce here as
\begin{equation}
p = \left[ \frac{1}{2\nu} \,   \frac{\nu^2  - B_0^2 P^2(2+ P)}{\nu^2 + B_0^2 P } - \nu\right] k^2 + \cdots  .
\label{eqphorizG0main}
\end{equation}
This gives the leading growth rate as a function of the parameters times $k^2$; it represents an eddy viscosity $\nu_E$ seen by large-scale modes and the instability is marked by this quantity becoming negative. While it is not possible to maximise this expression over $k$ to gain a complete analysis of the instability, it does give the instability threshold $\ReRe p = 0$, by setting $[\cdots] = 0$ to obtain 
\begin{equation}
B_0^2 = \frac{\nu}{P} \, \frac{ 1 - 2 \nu^2}{2\nu+ 2 P + P^2}\, .  
\label{eqG0threshold}
\end{equation}
This formula is plotted as the blue curve in panel \ref{fig-horiz-nu-B0}(b) and shows good agreement with the numerical results for the lower branch in panel  \ref{fig-horiz-nu-B0}(a). 
For $B_0=0$ we recover the hydrodynamic result $\nu_c = 1/\sqrt{2}$, and this analysis tells us how this hydrodynamic instability, domimated by the large-scale flow in $G_0$, is suppressed by interaction with the magnetic field. If we maximise $B_0^2$ as a function of $\nu$ in (\ref{eqG0threshold}), we find that this occurs at 
\begin{equation}
2 \nu^2=-Q + \sqrt{Q^2 +Q}, \quad Q = 2P + P^2, 
\label{eqG0thresholdB0}
\end{equation}
and putting this into $B_0^2$ gives an unwieldy expression for the threshold value $B_*$ above which the horizontal field suppresses the Kolmogorov instability. We do not present it here but give further discussion in \S6.

Note also that from (\ref{eqG0threshold}), 
\begin{equation}
B_0 \simeq  \frac{\nu}{P  \sqrt{2+P}} \, , \quad \nu \to 0,
\label{eqG0thresholdslope}
\end{equation}
and so the instability emerges with this slope from the origin $\nu =0$, $B_0=0$ of the figure. A typical example of the instability is shown in figure \ref{fig-horiz-G0-mode}: the stream function in panel (a) shows its nature as a zonostrophic instability giving horizontal jets, with modifications to the field in panel (b).

\begin{figure}
\centering
\hspace{-.5cm}  
(a)\includegraphics[scale=0.4]{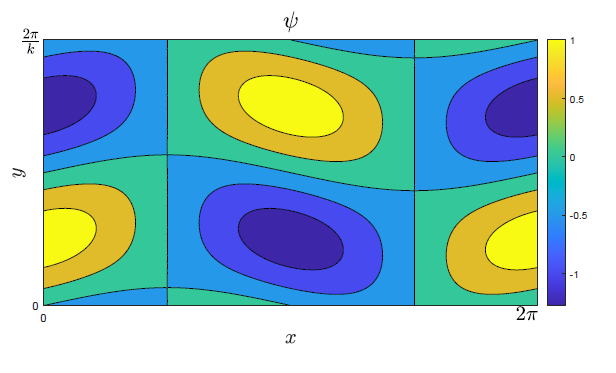}
(b)\includegraphics[scale=0.4]{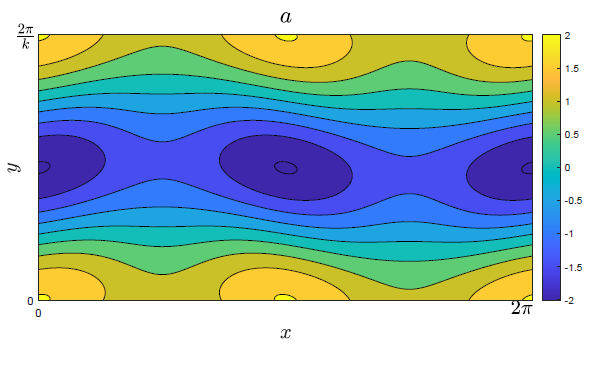}
 \caption{A typical unstable mode, with $B_0=0.25$, $\nu=\eta = 0.1$, $k=0.25$. from the field or $H_0$ branch; (a) shows the stream function $\psi$ and (b) the vector potential $a$.
 
 }
 \label{fig-horiz-H0-mode}
\end{figure}

The second system arising from perturbation theory involves $H_0$ and not $G_0$: it is dominated by a large-scale magnetic field and so we refer to this as the $H_0$ or \emph{field branch} of the horizontal field instability. The result of perturbation theory gives (\ref{eqphorizH0}), reproduced here as 
\begin{equation}
p = \left[  \frac{P}{2\nu} \, \frac{- \nu^2  +3 B_0^2 P }{\nu^2 + B_0^2 P } - \frac{\nu}{P} \right] k^2 + \cdots  .
\label{eqphorizH0main}
\end{equation}
The onset of instability again can be interpreted as a transport quantity becoming negative, here the eddy magnetic diffusivity $\eta_E$ \citep[see][]{chechkin1999negative}.  Note that this instability, being a directly growing instability does not connect to the strong-field branch of the vertical field (which is an over-stable wave).

The threshold for instability is given by $[\cdots] = 0$ or
\begin{equation}
B_0^2 = \frac{\nu^2}{P}\, \frac{ P^2 + 2 \nu^2}{3 P^2- 2 \nu^2}\, .  
\label{eqH0threshold}
\end{equation}
This curve is plotted on figure \ref{fig-horiz-nu-B0}(b) in red and again shows good agreement with the numerical results for the field branch in panel (a). Note that for fixed $P$, $B_0\to \infty$ as $\nu \to \nu_*$ with 
\begin{equation}
\nu_* = P \sqrt{3/2}, 
\label{eqhoriznustar}
\end{equation}
and so a viscosity larger than this is enough to prevent the field branch instability no matter how strong the field. We also have for small fields and viscosities that the threshold (\ref{eqH0threshold}) is given by 
\begin{equation}
B_0 \simeq  \frac{\nu}{\sqrt{3P}} \, , \quad \nu \to 0, 
\label{eqH0thresholdslope}
\end{equation}
and so for $P=1$ both thresholds (\ref{eqG0thresholdslope}) and (\ref{eqH0thresholdslope}) emerge from the origin with the same slope, though for general $P$ they are different. 
A typical example of the instability is given in figure \ref{fig-horiz-H0-mode}. The perturbation flow now does not have a zonostrophic jet structure, but shows closed eddies in panel (a). Panel (b) however shows a banded structure in the magnetic field (showing the dominant role of $H_0$), and indicates a tendency for the background mean field to segregate into bands of stronger and weaker horizontal field. 

\subsection{Instability for Bloch wave number $\ell\neq0$} 

Finally, we consider horizontal field for the case of non-zero $\ell$. This allows an instability to take up a scale $2\pi/k$ in the $y$-direction and $2\pi/\ell$, as $\ell \to 0$, in the $x$-direction. It turns out that instabilities can occur for $\ell \neq0$ even when the system is stable for $\ell=0$, in the case of horizontal field (unlike the situation for vertical field). 

\begin{figure}
\centering
\hspace{-.5cm}  
(a)\includegraphics[scale=0.367]{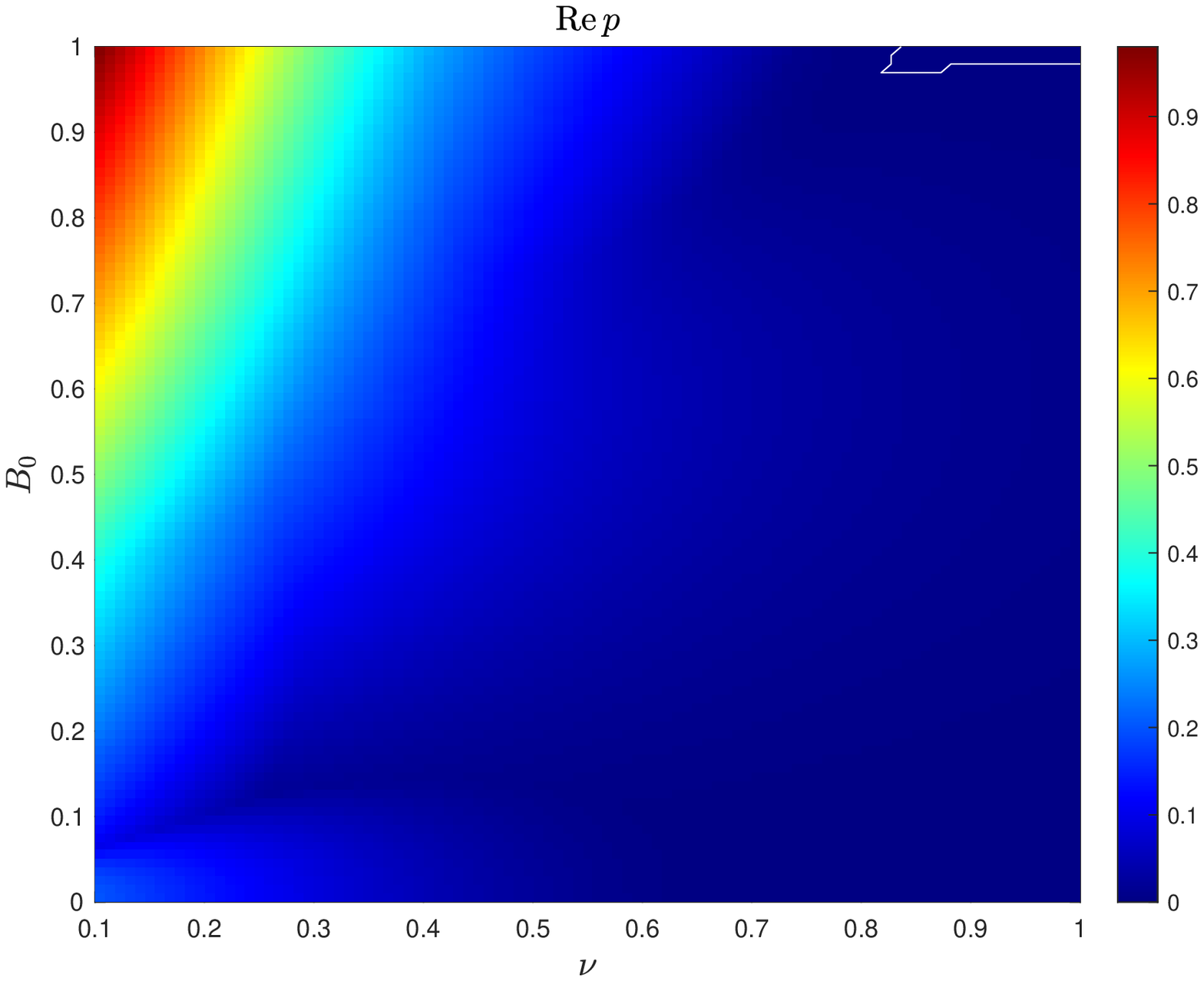}
(b)\includegraphics[scale=0.367]{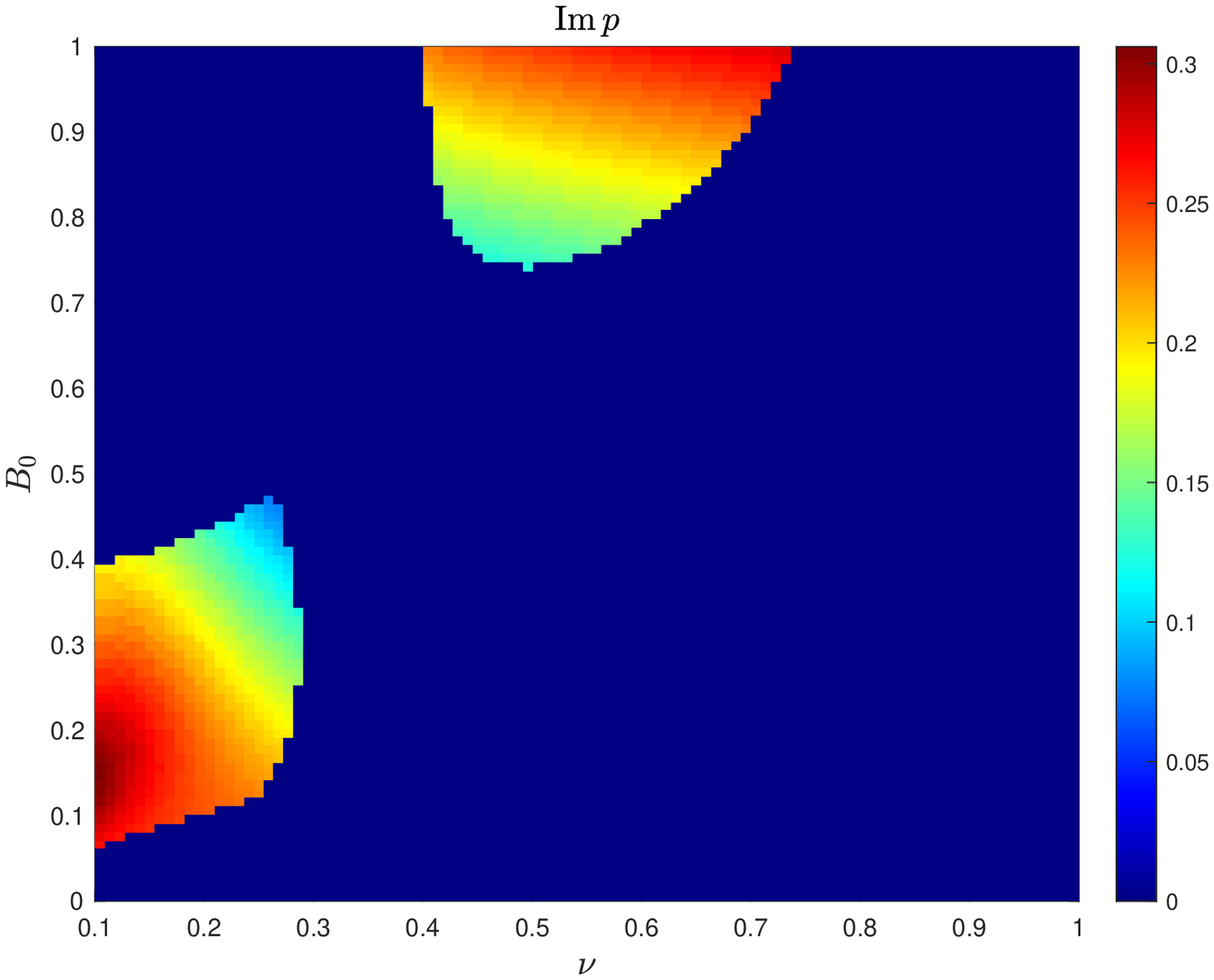}\\
(c)\includegraphics[scale=0.367]{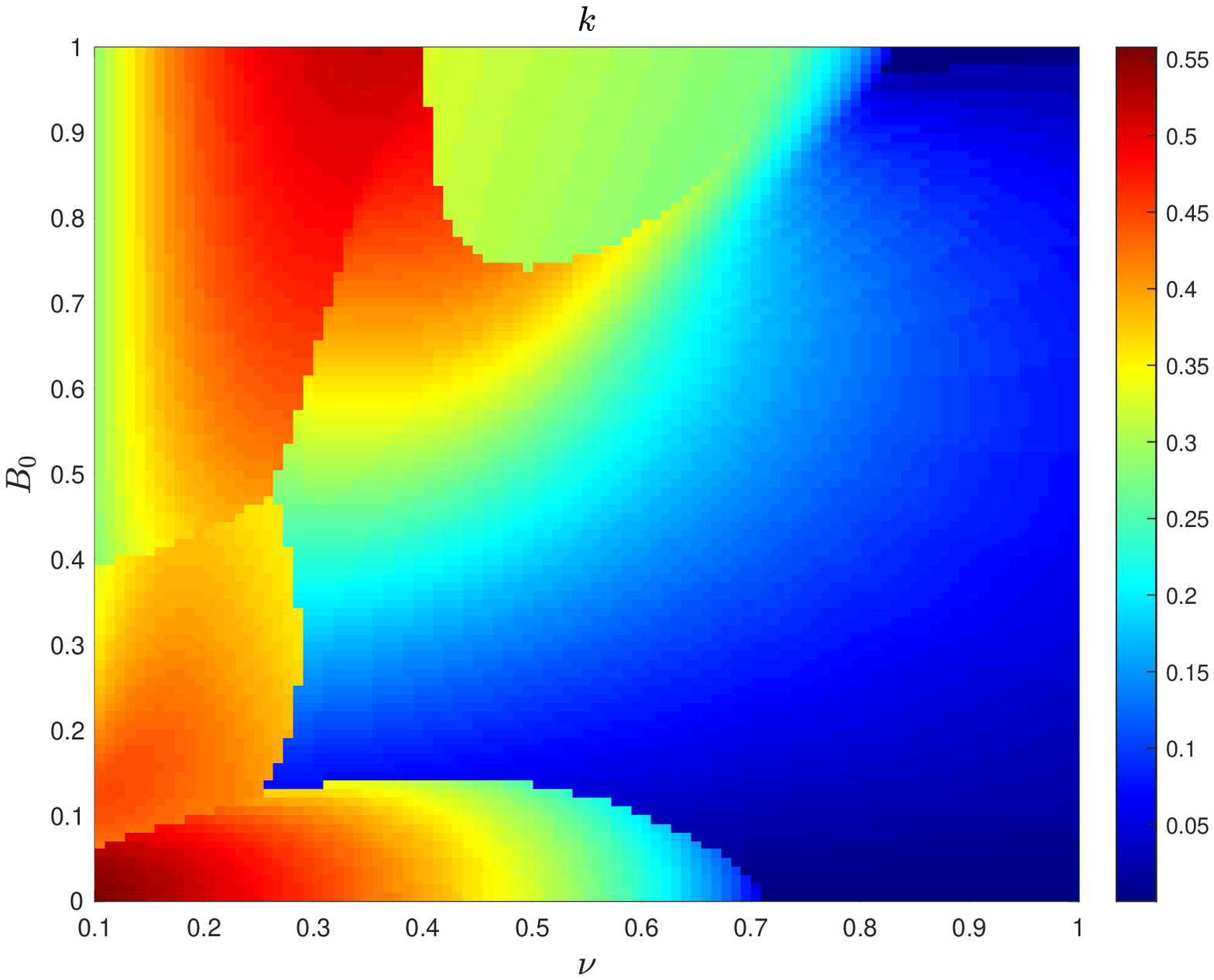}
(d)\includegraphics[scale=0.367]{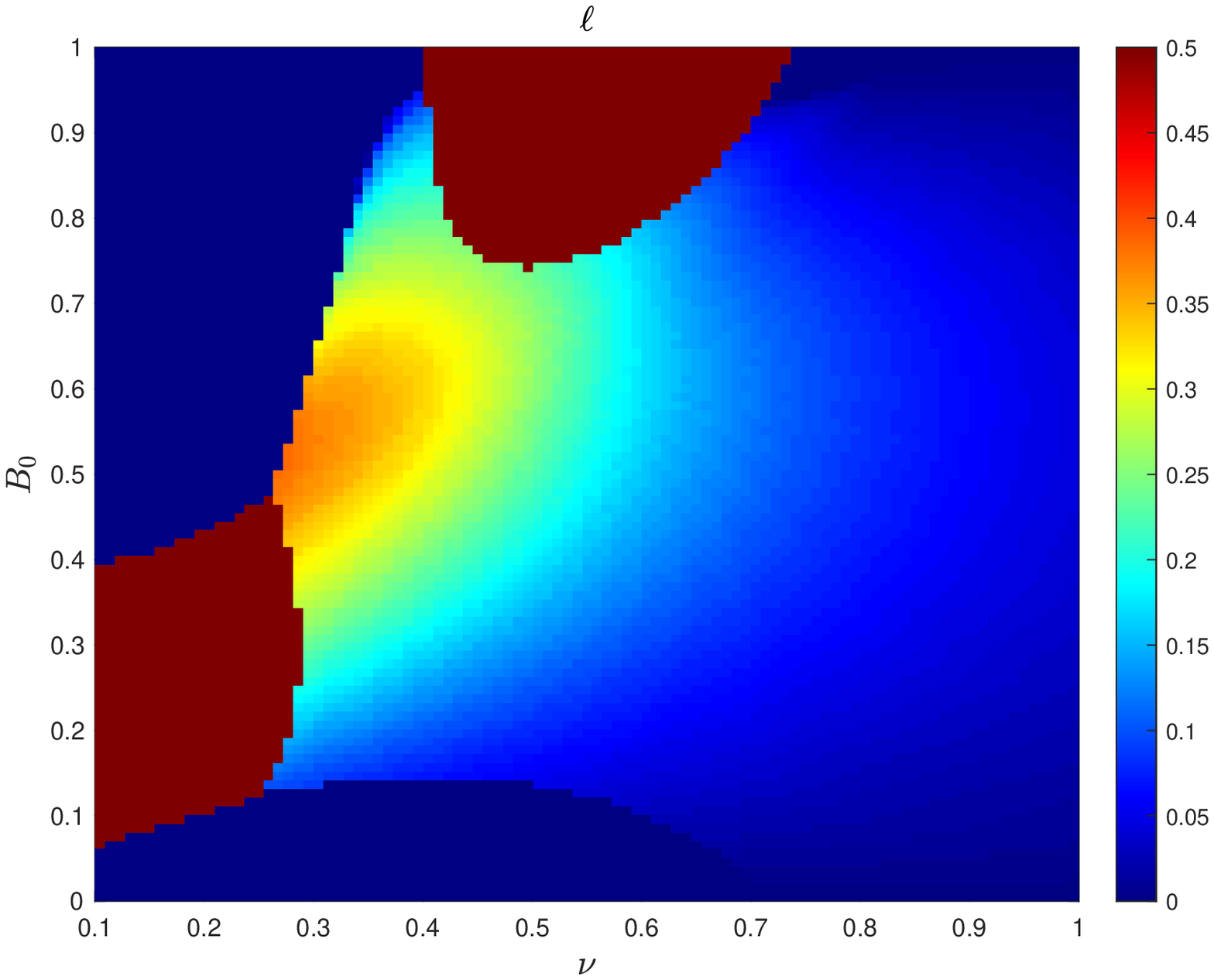}\\
\caption{(a) Instability growth rate $\ReRe p$ and (b) imaginary part $\ImIm p$ shown for horizontal field, plotted in the $(\nu, B_0)$ with $P=1$, $U_0=0$, and $\ell\neq0$. The maximising values of $k$ and of $\ell$ are shown in panels (c, d) respectively.
}
\label{fig-horiz-ell}
\end{figure}

Figure \ref{fig-horiz-ell}(a) shows a similar general structure to figure \ref{fig-horiz-nu-B0}(a), though note that the white curve has all but disappeared. The purely hydrodynamic instabilities are suppressed by increasing $B_0$, i.e.\ the $G_0$ or flow branch; this is not really visible now in panel (a) but the branch is evident in panel (c) showing $k_{\max}$, with $\ell_{\max}=0$ in panel (d) here. Then, the $H_0$ or field branch of instability appears clearly in panel (a) for larger $B_0$. However looking at panels (b) and (d) now the field branch has regions with non-zero imaginary part to the growth rate, regions with $\ell$ increasing from zero, and regions with $\ell$ constant at $\ell=0.5$ indicating instabilities with $4\pi$ periodicity in $x$ \cite[further investigated in][]{mephd2023}. Returning to figure  \ref{fig-horiz-ell}(a), the reason for the loss of the white curve, compared with figure \ref{fig-horiz-nu-B0}(a), is that there is now a new instability, reliant on having $\ell\neq0$ which means that almost all of the region of stability in the $\ell=0$ figure \ref{fig-horiz-nu-B0}(a) is now unstable in figure \ref{fig-horiz-ell}(a). The corresponding growth rates are quite small, and so this is otherwise not immediately evident on these plots.

\begin{figure}
\centering
\hspace{-.5cm}  
(a) \includegraphics[scale=0.5]{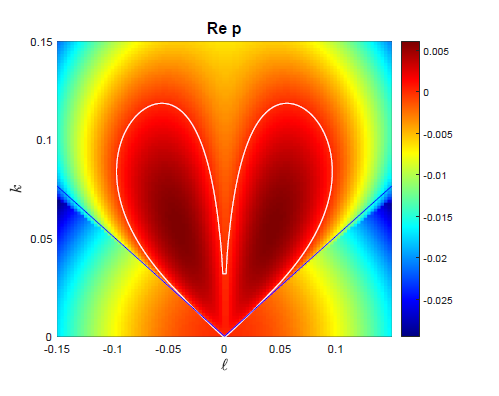}
(b) \includegraphics[scale=0.5]{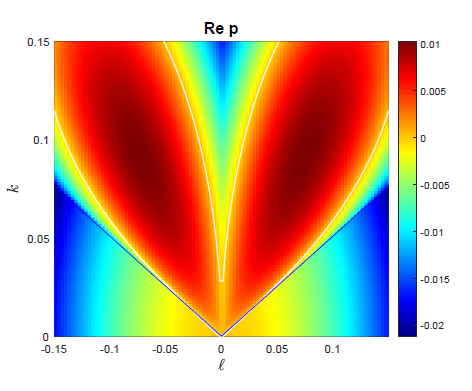}
 \caption{Instability growth rate $p$ for horizontal field as a function of $(\ell,k)$ for $\nu =\eta =0.75$ ($P=1$) with $B_0 = 0.2$, (a) numerical growth rates and (b) approximate growth rates calculated from (\ref{eqhorizellpfixed}). 

 }
 \label{fig-horiz-butterfly}
\end{figure}

\begin{figure}
\centering
\hspace{-.5cm}  
(a)\includegraphics[scale=0.4]{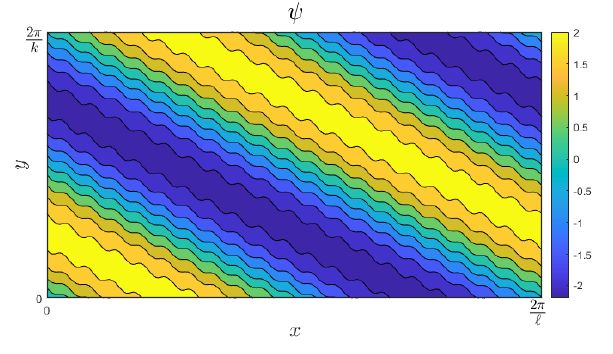}
(b)\includegraphics[scale=0.4]{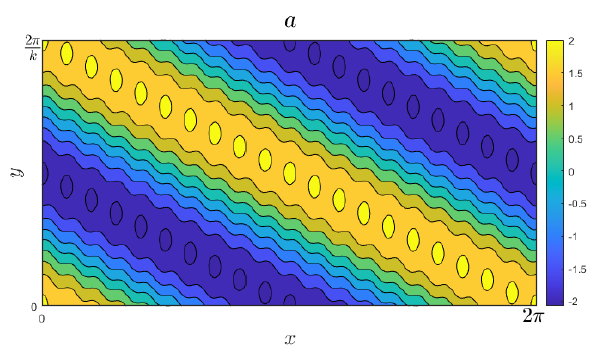}
 \caption{A typical unstable mode, with $\ell = 0.05$, $k = 0.05$, $\nu=\eta = 0.75$, $B_0 = 0.2$; (a) shows the stream function $\psi$ and (b) the vector potential $a$.
  
 }
 \label{fig-horiz-ell-mode}
\end{figure}

We therefore turn to theory for $\ell\neq0$, developed in appendix D, which gives a growth rate in (\ref{eqhorizellp}) of 
%
\begin{equation}
p =   \pm B_0 \ell \left[ \frac{k^2}{\ell^2+k^2} \, \frac{ P [ \nu^2 (P+2) - P^2 B_0^2] }{\nu^2 (\nu^2 + PB_0^2)} -  1 \right]^{1/2} + \cdots . 
\label{eqhorizellpmain}
\end{equation}
This approximation reveals an instability that crucially relies on having a non-zero Bloch wavenumber, $\ell\neq0$, with $\ell$ and $k$ both small. 
If we fix the parameters $\nu$, $P$ and $B_0$ we can consider the growth rate $p$ as a function in the $(\ell,k)$ plane. Setting the quantity inside the square root to zero to find a threshold, we see that the region of instability is demarcated by the pair of straight lines given by
\begin{equation}
\frac{k^2}{\ell^2} = 
\frac{\nu^2 (\nu^2 + P B_0^2)}{\nu^2 (P^2 + 2P - \nu^2) - PB_0^2(\nu^2+P^2)}\, . 
\label{eqhorizstraightlines}
\end{equation}

The growth rate $p$ tells us about the instability growth rate as we increase $k$ and $\ell$ from zero, but to find how this eventually decreases, we would need to go to next order in perturbation theory, which is impractical and unlikely to be informative. To give a qualitative feel for the growth rate we will add on the diffusive suppression term $ - \tfrac{1}{2} (\nu + \eta) k^2$ that is certainly present at next order, and look at 
\begin{equation}
p =   \pm B_0 \ell \left[ \frac{k^2}{\ell^2+k^2} \, \frac{ P [ \nu^2 (P+2) - P^2 B_0^2] }{\nu^2 (\nu^2 + PB_0^2)} -  1 \right]^{1/2} - \tfrac{1}{2} \nu (1 + P^{-1}) ( k^2 +\ell^2) + O(k^2, \ell^2) . 
\label{eqhorizellpfixed}
\end{equation}
To see how this works, figure \ref{fig-horiz-butterfly}(a) shows growth rates plotted in the $(\ell,k)$-plane for $B_0 =  0.2$ and $\nu = \eta = 0.75$, parameters corresponding to stability for $\ell=0$. There is now a region of instability taking a `butterfly' form, outlined by the white curve $\ReRe p = 0$. The straight blue lines are given by (\ref{eqhorizstraightlines}) and are tangential to the white lines at the origin, confirming the theory. In panel (b) we show an analogous figure for the `fixed up' growth rate in (\ref{eqhorizellpfixed}). The agreement between the results in the two panels (a) and (b) is excellent near the origin, but then further out the agreement is only qualitative, and quite rough,  as we might expect. For an example of an unstable configuration, figure \ref{fig-horiz-ell-mode} shows the flow and field for the fastest growing mode in figure \ref{fig-horiz-butterfly}. We observe the structure of a large-scale jet-like structure but at an oblique angle to the $y$-direction, this being allowed by the non-zero  Bloch wavenumber $\ell$. 

Returning to the bigger picture, for instability at a  general points in the $(\nu, B_0)$ plane we need the quantity $[\cdots]$ inside the square root in (\ref{eqhorizellpmain}) to be positive for some values of $k$ and $\ell$. Equivalently, it corresponds to requiring that the straight lines in (\ref{eqhorizstraightlines}) have a finite slope.  It can be checked that this gives a threshold for instability:  
\begin{equation}
B_0^2 
= \frac{\nu^2 (P^2 + 2P - \nu^2)}{P(\nu^2 + P^2)}\, . 
\label{eqhorizB0ellthreshold} 
\end{equation}
Values of field smaller than this give instability and so this formula gives the threshold curve in the $(\nu,B_0)$ plane for this family of $\ell\neq0$ instabilities.  This threshold is shown as a dashed curve in figure \ref{fig-horiz-nu-B0}(b). We thus see in this panel that allowing $\ell\neq0$ means that almost the whole of the parameter ranges shown give instability, all except for a small curved triangular region at the top right, and this is in agreement with the white curve obtained numerically in figure \ref{fig-horiz-ell}(a). The agreement is not perfect because the growth rates in the top right corner become small and the unstable region shrinks away in the $(\ell,k)$-plane, as the thresholds are approached, and so the precise location of the white curve becomes hard to resolve without further work. 

Note that for $P=1$, from (\ref{eqhorizB0ellthreshold})  the instability is cut off at $\nu = \sqrt{3}$, and in general the instability requires
\begin{equation}
\nu < \nu_* = \sqrt{P(2+P)}. 
\label{eqhorizB0nustar1} 
\end{equation}
Numerical experimentation shows that for $\nu$ less than this value, the region of instability in the $(\ell,k)$-plane becomes vanishingly small as $B_0$ tends to zero or tends to the value given in (\ref{eqhorizB0ellthreshold}). The maximum magnetic field for the $\ell\neq0$ instability found here is given by maximising $B_0$ over $\nu$: the maximum occurs at 
\begin{equation}
\nu^2 = - P^2 + \sqrt{2P^3(1+P)} , 
\label{eqhorizB0numax} 
\end{equation}
and we pick this up in the final discussion section, \S6.


\section{Discussion}

In this study we have explored instability of the classic Kolmogorov flow in the presence of magnetic field which is either vertical, aligned with the flow, or horizontal, aligned with possible jet formation. In the vertical field case we have obtained new analytical results for the maximum growth rate (\ref{eqappCpmax}) and magnetic field threshold (\ref{eqappCpthreshold}), that show the suppression of the original instability found by \cite{meshalkin1961investigation} by vertical magnetic field. In particular we have obtained a value $B_*$ in (\ref{eqvertweakfieldbstar}) as an estimate of the field magnitude required to suppress the Kolmogorov instability. This corresponds to a threshold that is   
\begin{equation}
B_*^2 \sim P^{-1} = \frac{\eta}{\nu} \quad (P\ll 1), \qquad
B_*^2 \sim P = \frac{\nu}{\eta}\ \quad (P\gg 1) . 
\end{equation}
For the strong field branch we have confirmed the numerical results of \cite{fraser2022non} and provided an alternative derivation of their growth rate formula (\ref{eqstrongvertfieldRep0}) and threshold (\ref{eqstrongvertfieldRep1}). We note that these authors also find a third branch of instabilities, which they term `varicose Kelvin--Helmholtz' modes, which we have not observed at the Reynolds numbers we have used.

The case of horizontal field, broadly relevant to several studies of jet formation where the field is aligned with potential jets \citep{tobias2007beta,durston2016transport,constantinou2018magnetic}, shows more complex structure, unsurprisingly given the wavey nature of the magnetic field in the basic state, seen in figure \ref{fig-basic-state}(b). We observe again the suppression of the purely hydrodynamic instability, the flow or $G_0$ branch, when the magnetic field strength is increased, with a threshold of $B_0= B_*$ for complete suppression with $B_*$ given by substituting $\nu^2$ from (\ref{eqG0thresholdB0}) into  (\ref{eqG0threshold}). For large $P$, we have $\nu^2 \simeq 1/4$ while for small $P$, $\nu^2 \simeq\sqrt{P/2}$. The value of $B_*$ for suppression then amounts to:
\begin{equation}
B_*^2 \sim P^{-1} = \frac{\eta}{\nu}\ \quad (P\ll 1), \qquad 
B_*^2 \sim P^{-3} = \frac{\eta^3}{\nu^3} \quad (P\gg 1). 
\end{equation}
Interestingly this bears comparison with \cite{tobias2007beta} who have $\nu = 10^{-4}$ fixed and $10^{-1} \leq \eta \leq 10^{-6}$ in their runs. For the greater values of $\eta$ used, $P\ll1$ and so a threshold $B_*^2 \sim \eta$ is indicated above, and found in these full numerical simulations. Also, note that at this threshold we have that the forcing magnitude is fixed in magnitude in (\ref{eqhorizbasicstate}) and so there is, at least roughly, a correspondence of working with fixed forcing amplitude as in their paper and the fixed Kolmogorov flow in ours. However we should remark that these authors use a non-zero value of $\beta$ and we have $\ell$ zero, and so further work would need to be done to make a sound comparison. 

A feature of the horizontal field problem is that  for increasing magnetic field strengths a further branch of instabilities emerges, the field or $H_0$ branch, also seen in \cite{durston2016transport}. Analytical formulae for the  threshold of instability are given for each branch. The field branch exists provided the Reynolds number is above a threshold, $\Rey > \nu_*^{-1}$ with $\nu_*$ given in (\ref{eqhoriznustar}). Allowing a Bloch wavenumber $\ell\neq0$ in the $x$-direction, in addition to the wavenumber $k$ in the $y$-direction, allows a new branch of instabilities. In particular a magnetic field, no matter how weak, can destabilise the Kolmogorov flow provided the  Reynolds number $\Rey=\nu^{-1} > \nu_*^{-1} $ with $\nu_*$ given in (\ref{eqhorizB0nustar1}). For example at $P=1$, the purely hydrodynamic instability is present for $\Rey > \sqrt{2}$ but an MHD instability is present for arbitrarily weak but non-zero horizontal magnetic field provided $\Rey > 1/\sqrt{3}$. For sufficiently large magnetic field this instability is again suppressed, and making use of (\ref{eqhorizB0numax}) (with $\nu^2 \simeq \sqrt{2P^3}$ for small $P$ and $\nu \simeq (\sqrt{2}-1){P^2}$ for large $P$) we find a threshold 
\begin{equation}
B_*^2 \sim 1  \quad (P\ll 1), \qquad 
B_*^2 \sim P = \frac{\nu}{\eta} \quad (P\gg 1). 
\end{equation}
Note that the order of limits could be important: in our discussion in this paper we are fixing any value of $P$ and then allowing $k$ and $\ell$ to tend to zero. Other limits are possible and could be explored by appropriate scalings in our calculations. 

Underlying our study is matrix eigenvalue perturbation theory as set out in the appendices, a flexible tool for these types of problems. We find it gives greater clarity than using a multiple scales formulation or applying perturbation theory to roots of a polynomial, even though all these methods are ultimately equivalent. Note that while many of the instabilities seen by us and by other authors can be characterised as involving a negative eddy viscosity term  $- \nu_E k^2$ or a negative eddy magnetic diffusivity term $- \eta_E k^2$ at large scales, the growth rate $p(k,\ell)$ in the case of horizontal field shows a complicated dependence on $k$ and $\ell$ in (\ref{eqhorizellpmain}). Although this $\ell\neq0$ instability occurs at arbitrarily large scales, it cannot be categorised as involving a simple negative eddy transport effect. This arises because we are applying perturbation theory to a repeated eigenvalue of the limiting $k\to0$, $\ell\to0$ problem. Looking to the future, it would be interesting to pursue further research on the Kolmogorov flow as an MHD system, particularly on the nonlinear evolution of instabilities and inverse cascades \citep{fraser2022non, mephd2023}, and on the interaction of magnetic field with a $\beta$-effect and Rossby waves.

\appendix



\section{Horizontal field, with $U_0=0$, $\ell=0$}

First, let us outline the general principles of our approximate analysis of growth rates in this and other appendices. We have in each case an infinite  system of coupled linear equations, for example (\ref{eqGnhoriz}, \ref{eqHnhoriz}) in the case of horizontal field with $U_0=0$ and $\ell=0$. In the limit $k \to 0$ of large-scale modes, we can reduce the calculation of the growth rate to eigenvalues of a finite matrix. The key point to note is that the $n=0$ modes $G_0$ and $H_0$ are distinguished from all other modes by being weakly damped, as $p_{\mathrm{visc}}  = -\nu k^2$ or $p_{\mathrm{diff}}  = -\eta k^2 $ by molecular diffusion with $k\ll1$. On the other hand other modes with $|n|>0$ are relatively strongly damped, with $p_{\mathrm{visc}}  = -\nu (n^2+ k^2)$ or $p_{\mathrm{diff}}  = -\eta (n^2+k^2) $. So we need to keep the $n=0$ modes as these are most easily destabilised in the system. How are they destabilised? This is through the coupling from $n=0$ to low $n\neq0$ modes and then back to $n=0$, and via these couplings the unstable mode can draw energy from the basic state Kolmogorov flow $\uv_0 = (0, \sin x)$. It then follows that we typically expect to see effects only at second order or beyond in terms of perturbation theory, and this then will involve $G_n$ and $H_n$ for $n = -1$, $0$, $1$ only. Thus we can truncate the system and then use perturbation theory for the eigenvalues of a finite matrix, with $k\ll1$ as an expansion parameter.  

Thus we set to work on the horizontal system of equations (\ref{eqGnhoriz}, \ref{eqHnhoriz}) in Fourier space. These are truncated to just the modes $G_0$, $G_{\pm 1}$, $H_0$ and $H_{\pm 1}$, with  
\begin{align}
pG_0  & = -\nu k^2 G_0
 -\frac{k}{2} \frac{k^2}{1+k^2} \, G_{-1} 
 +\frac{k}{2} \frac{k^2}{1+k^2} \, G_{1} 
  + \frac{ikB_0}{2\eta} \, k^2 H_{-1} 
+ \frac{ikB_0}{2\eta} \,  k^2 H_{1} , 
\label{eqG0vert}
\\
pH_0 & =- \eta k^2 H_0   -\frac{k}{2}\, H_{-1}+\frac{k}{2}\, H_{1} 
 + \frac{ikB_0}{2\eta}  \frac{1}{ 1 + k^2} \, G_{-1} 
+ \frac{ikB_0}{2\eta}  \frac{1}{ 1 + k^2} \, G_{1} , 
\\
pG_{\pm1}  & =  -\nu(1+k^2) G_{\pm1} \pm \frac{k}{2} \, \frac{1-k^2}{k^2}\, G_{0}  
  \pm i B_0(1 + k^2)H_{\pm1} 
+ \frac{ikB_0}{2\eta} (  -1  + k^2) H_{0} , 
 \label{eqGpmvert}
\\
pH_{\pm1} & =- \eta(1+k^2)H_{\pm 1}  \mp \frac{k}{2}\, H_{0} 
\pm  i B_0\,  \frac{1}{1+k^2} \, G_{\pm 1} 
  + \frac{ikB_0}{2\eta}  \frac{1}{k^2} \, G_{0} .
\end{align}
We now express these equations in terms of $G_0$, $H_0$ and the fields
\begin{equation}
G_{\pm} = \half ( G_{1} \pm G_{-1}), \quad 
H_{\pm} = \half ( H_{1} \pm H_{-1}) ; 
\label{eqGpmHpmdef} 
\end{equation}
we rescale $G_0 = G'_0 k^2$ and for convenience we set $\Bt_0 = B_0 / \eta$. The resulting equations then break up into two uncoupled systems. The first involves only $G'_0$ on the large scale,
\begin{align}
pG'_0  & = -\nu k^2 G'_0
 + {k}(1+k^2)^{-1} G_{-} 
  + ik\Bt_0 H_+,
\label{eqG0horiz1}
\\
pG_-  & =  -\nu(1+k^2) G_- + \half k (1 - k^2 )G'_{0}  
  +  i B_0(1 + k^2)H_+ ,
 \label{eqG0horizt2}
 \\
pH_+ & =- \eta(1+k^2)H_{+}  
+ iB_0 (1+k^2)^{-1} G_-
  + \half ik\Bt_0 G'_0,
   \label{eqG0horiz3}
 \end{align}
while the second involves only $H_0$ on the large scale,
\begin{align}
pH_0 & =- \eta k^2 H_0  
 + k H_{-} 
 + ik\Bt_0\,   ( 1 + k^2)^{-1} \, G_{+} ,
\label{eqH0horiz1}
 \\
pH_- & =- \eta(1+k^2)H_{-}  
-\half k  H_{0} 
+ iB_0 (1+k^2)^{-1} G_+ , 
\label{eqH0horiz3}
\\
pG_+  & =  -\nu(1+k^2) G_+ 
  +i B_0(1 + k^2)H_- 
+ \half {ik\Bt_0} (  -1  + k^2) H_{0} ,
\label{eqH0horiz2}
 \end{align}
We deal with these two branches in turn, using eigenvalue perturbation theory. 

\subsection{Outline of approach} 

Having reduced the problem to the calculation of eigenvalues $p$ for two systems of equations for $k\ll1$, one for (\ref{eqG0horiz1}--\ref{eqG0horiz3}) and one for (\ref{eqH0horiz1}--\ref{eqH0horiz3}), we outline the method that is common to all the approximations in the paper. In each case we write the governing eigenvalue problem in a matrix form (see (\ref{eqvffullM}, \ref{eqvffullMH0}) below) where the matrix $M$ depends on $k$, $\vv$ is the eigenvector and $p$ the eigenvalue: 
\begin{equation}
M\vv = p \vv. 
\end{equation}
We may rescale some quantities in the matrix $M$ (using a prime to denote these) and we then proceed to expand $M$ in powers of $k$ as 
\begin{equation}
M = M_0 + k M_1 + k^2 M_2 + \cdots, 
\end{equation}
and likewise $\vv$ and $p$. For the limit $k \to 0$ we solve 
\begin{equation}
(M_0 + k M_1 + \cdots) ( \vv_0 + k \vv_1  + \cdots) = (p_0 + kp_1 + \cdots)  (\vv_0 + k \vv_1 + \cdots), 
\label{eqbasicM0series}
\end{equation}
order by order in $k$. Here we set out the first few orders in a convenient form: 
\begin{align}
 p_0 \vv_0 & = M_0 \vv_0 ,  
 \label{eqgenprob0}\\
 p_1 \vv_0 & = (M_0 - p_0) \vv_1 + M_1 \vv_0, 
  \label{eqgenprob1}\\
 p_2 \vv_0 & = (M_0 - p_0) \vv_2 + M_2 \vv_0 + M_1 \vv_1 - p_1 \vv_1 , 
  \label{eqgenprob2} \\
 p_3 \vv_0 & = (M_0 - p_0) \vv_3 + M_3 \vv_0 + M_2 \vv_1 + M_1 \vv_2 - p_2 \vv_1 - p_1 \vv_2, 
  \label{eqgenprob3}\\
 p_4 \vv_0 & = (M_0 - p_0) \vv_4 + M_4 \vv_0 + M_3 \vv_1 + M_2 \vv_2 + M_1 \vv_3 - p_3 \vv_1 - p_2 \vv_2 - p_1 \vv_3 .
  \label{eqgenprob4}
 \end{align}
First we choose an eigenvalue $p_0$ and corresponding eigenvector $\vv_0$ of $M_0$; at the level of $M_0$ the mode is undamped and so the real part of $p_0$ is zero. Assuming this is a simple (non-repeated eigenvalue) there is also a single left eigenvector $\wv_0$ with $\wv_0 (M_0 - p_0) = 0$. With (\ref{eqgenprob0}) thus dealt with, we note that we gain successive eigenvalues from applying $\wv_0$ to the left of the remaining equations, so that
\begin{align}  
 p_1 \, \wv_0 \vv_0 & = \wv_0 M_1 \vv_0, 
  \label{eqgenprob1a}\\
 p_2\,  \wv_0  \vv_0 & =    \wv_0 ( M_2 \vv_0 +   M_1 \vv_1 - p_1 \vv_1 ), 
  \label{eqgenprob2a}
  \\
p_3\,  \wv_0 \vv_0 & =    \wv_0 ( M_3 \vv_0 + M_2 \vv_1 + M_1 \vv_2 - p_2 \vv_1 - p_1 \vv_2) , 
  \label{eqgenprob3a}\\
 p_4\,  \wv_0 \vv_0 & =  \wv_0( M_4 \vv_0 + M_3 \vv_1 + M_2 \vv_2 + M_1 \vv_3 - p_3 \vv_1 - p_2 \vv_2 - p_1 \vv_3 ). 
   \label{eqgenprob4a}
 \end{align}
In this way once having chosen the eigenvalue $p_0$ to perturb from (\ref{eqgenprob0}) together with $\vv_0$ and $\wv_0$, we find $p_1$ from (\ref{eqgenprob1a}). We then need $\vv_1$ from (\ref{eqgenprob1}) and while $M_0- p_0$ is not invertible, having fixed the value of $p_1$, there is a solution for $\vv_1$. It is not unique, but this does not matter as we shall see. We can then calculate $p_2$ from (\ref{eqgenprob2a}) and so forth. We will go up to the level of $p_4$ in some of our calculations below. 

\subsection{Flow or $G_0$ branch} 

Having set out our general approach we return to the first system  (\ref{eqG0horiz1}--\ref{eqG0horiz3}), which involves a dominant large-scale flow in $G_0$ and no large-scale field, with 
\begin{equation}
M = 
\begin{pmatrix} 
- \nu k^2  & k(1+k^2)^{-1} & ik\Bt_0 \\ 
\;\half k (1-k^2)\;  & \;-\nu (1+k^2)\; & \;iB_0 (1+k^2)\; \\ 
\half ik \Bt_0  & iB_0 (1+k^2)^{-1}  & -\eta (1+k^2)  \\ 
\end{pmatrix} , 
\quad
\vv = \begin{pmatrix} G'_0 \\ G_- \\ H_+ \end{pmatrix} .
\label{eqvffullM}
\end{equation}
For this branch we expand $M$ to give:  
\begin{equation}
M_0 = 
\begin{pmatrix} 
0 & 0 &  0 \\
0 & -\nu &  \;iB_0 \;\\
\;0\; &\; iB_0\; &  -\eta 
\end{pmatrix}, \quad
M_1 = 
\begin{pmatrix} 
0 & \;1\; &  \;i\Bt_0\; \\
\half & 0 &  0 \\[1pt]
\; \half i\Bt_0\; & 0 &  0\
\end{pmatrix}, \quad
M_2 = 
\begin{pmatrix} 
\;-\nu\; & 0 &  0 \\
0 & \;-\nu\; & \; iB_0\; \\
0 & -iB_0 & -\eta 
\end{pmatrix}.
\end{equation}
Note that the inverse of the non-trivial $2\times 2$ block of $M_0$ is 
\begin{equation}
\begin{pmatrix} 
\;-\nu \;&  iB_0 \\
 iB_0 & \; -\eta\; \\
\end{pmatrix}^{-1} = 
\Delta 
\begin{pmatrix} 
\;-\eta\; &  -iB_0 \\
- iB_0 &  \;-\nu\; 
\end{pmatrix}, 
\quad
\Delta^{-1}  =\nu \eta   + B_0^2, 
\label{eqhfDeltadef0}
\end{equation}
where $\Delta$ is the inverse of the appropriate determinant.

We are ready to solve order by order in $k$. At leading order (\ref{eqgenprob0}) we choose
\begin{equation}
p_0 = 0 , \quad \vv_0 = (1,0,0)^T, \quad \wv_0 = (1,0,0) .
\label{eqlopvwsolution}
\end{equation}
At next order, we use (\ref{eqgenprob1a}) and have 
\begin{equation}
M_1 \vv_0 = (0, \half , \half i \Bt_0)^T, \quad 
p_1 = 0 . 
\end{equation}
 Given this we now solve (\ref{eqgenprob1}) for $\vv_1$, to obtain
\begin{equation}
\vv_1 =  \half \Delta (0, \eta - \Bt_0 B_0 , iB_0 + i \nu \Bt_0)^T .
\end{equation}
Here we have used the inverse (\ref{eqhfDeltadef0}) of the $2\times2$ block of $M_0$ to find a solution for $\vv_1$. We could add on an arbitrary multiple of $\vv_0$ to this, but this would only change the (irrelevant) normalisation of the eigenvector $\vv$ in our calculation --- any solution is acceptable.

Finally at $O(k^2)$ we find from (\ref{eqgenprob2a}) the value of $p_2$ and, recalling that $\Bt_0 = B_0/\eta$, this gives
\begin{equation}
p = p_2 k^2 + \cdots  =\bigl[  \half \Delta ( \eta - 2 B_0^2 / \eta - B_0^2 \nu / \eta^2) - \nu\bigr] k^2 + \cdots  ,
\end{equation}
which is, with the Prandtl number $P = \nu/ \eta$,
\begin{equation}
p = \left( \frac{1}{2\nu} \,   \frac{\nu^2  - B_0^2 P^2 (2+P) }{\nu^2 + B_0^2 P } - \nu\right) k^2 + \cdots  .
\label{eqphorizG0}
\end{equation}
We pick up the discussion in the main part of the paper, at (\ref{eqphorizG0main}). 
%


\subsection{Field or $H_0$ branch}

In the second system (\ref{eqH0horiz1}--\ref{eqH0horiz3}), the large-scale field $H_0$ is present but no large-scale flow. We write the system as $M\vv = p\vv$ with 
\begin{equation}
M 
= 
\begin{pmatrix} 
- \eta k^2 & k & ik\Bt_0 (1+k^2)^{-1}  \\ 
- \half k  & - \eta (1+k^2) & iB_0 (1+k^2)^{-1}  \\ 
\;\half ik \Bt_0 (-1 + k^2)  \;& \;iB_0 (1+k^2) \;& \;-\nu (1+k^2) \; \\ 
\end{pmatrix} , \quad
\vv = 
\begin{pmatrix} H_0 \\ H_- \\ G_+ \end{pmatrix} .
\label{eqvffullMH0}
\end{equation}
The matrix series for $M$ now has 
\begin{equation}
M_0 = 
\begin{pmatrix} 
0 & 0 &  0 \\
0 & -\eta &  iB_0 \\
\;0\; & \;iB_0 \;& \;-\nu\;  
\\
\end{pmatrix}, \quad
M_1 = 
\begin{pmatrix} 
0 & \;1\; &  \;i\Bt_0\; \\
-\half & 0 &  0 \\
\;- \half i\Bt_0 \;& 0 &  0 \\
\end{pmatrix}, \quad
M_2 = 
\begin{pmatrix} 
\;-\eta\; & 0 &  0 \\
0 & \;-\eta\; &  -iB_0 \\
0 & iB_0 &  \;-\nu\; \\
\end{pmatrix}.
\end{equation}
We have that the inverse of the $2\times 2$ block of $M_0$ is given as in (\ref{eqhfDeltadef0}) with $\nu$ and $\eta$ interchanged and the same $\Delta$. The calculation proceeds as before. At leading order in the eigenvalue problem   (\ref{eqbasicM0series}) we take the same solution as that given in (\ref{eqlopvwsolution}). At first order, we have 
\begin{equation}
M_1 \vv_0 = (0, - \half , - \half i \Bt_0)^T.
\end{equation}
This gives $p_1 = 0 $ and we solve (\ref{eqgenprob1}) for  $\vv_1$ as
\begin{equation}
\vv_1 =  \half \Delta (0, - \nu  + \Bt_0 B_0 , -2iB_0 )^T . 
\end{equation}
At the next order (\ref{eqgenprob2a})  yields $p_2$ and so 
\begin{equation}
p = p_2 k^2 + \cdots  =\bigl[  \half \Delta ( -\nu + 3B_0^2 / \eta  ) - \eta\bigr] k^2 + \cdots  , 
\end{equation}
or, with $P = \nu/ \eta$,
\begin{equation}
p = \left(  \frac{P}{2\nu} \, \frac{- \nu^2  +3 B_0^2 P }{\nu^2 + B_0^2 P } - \frac{\nu}{P} \right) k^2 + \cdots  .
\label{eqphorizH0}
\end{equation}
Further analysis commmences from equation (\ref{eqphorizH0main}). 




\section{Vertical strong field, with $U_0 = 0$, $\ell=0$}

In this appendix we turn to the vertical field system. Here there are two types of instability and two analyses that we will set out in this appendix and the next one. The calculation in this appendix is designed to capture the properties of the strong field branch seen for $\eta>\nu$ in figure \ref{fig-strong-vertical-nu-B0} and is equivalent to that set out in \cite{fraser2022non}. Mathematically we need to consider the limit when $B_0 \to\infty $ as $k\to 0$, and we find that relating these via $B_0 = O(k^{-1})$ is most informative. We will  set the Bloch wavenumber $\ell = 0$ and take no mean flow $U_0=0$.

If we write out the vertical field equations truncated to $G_0$, $H_0$, $G_{\pm 1}$ and $H_{\pm 1}$, rewrite in terms of $G_{\pm}$ and $H_{\pm}$ defined in (\ref{eqGpmHpmdef}), then we obtain the equations (without any further approximation) in the form $M\vv = p\vv$ with 
\begin{equation}
M = \left(
\begin{array}{cc|cc}
- \nu k^2 &  ikB_0   & k(1+k^2)^{-1}&0       \\
    ikB_0 & -\eta k^2& 0  &k        \\
  \hline
    \half  k(1-k^2)  &   0 & -\nu (1+k^2) & ikB_0 (1+k^2)  \\ 
0 &  -\half  k &  ikB_0 (1+k^2)^{-1}  & - \eta (1+k^2)      \\  
 \end{array}
\right), \quad
\vv =  \begin{pmatrix} G'_{0} \\ H_0 \\ G_{-} \\ H_{-}  \end{pmatrix}, 
\label{eqfullvertfieldmatrix}
\end{equation}
where $G_0 = k^2 G'_0$ as usual. The fields $G_{+}$ and $H_{+}$ are decoupled (as we have $U_0=0$) and so not considered further. Before expanding $M$ in powers of $k$, for strong vertical field we  rescale $B_0 = k^{-1} B'_0$ with $B'_0$ fixed in the limit $k \to 0$. Then expanding $M$ gives the matrices
\begin{equation}
M_0 = \left(
\begin{array}{cc|cc}

0 &\; iB'_0 \;& 0 & 0 \\
\;iB'_0\; & 0 & 0 & 0 \\
   \hline
0 & 0 & \;- \nu\; & \;iB'_0\; \\
0 & 0 & iB'_0 & - \eta \\
  \end{array}
\right), 
\quad
M_1 = \left(
\begin{array}{cc|cc}
\; 0\; & \;0\;  & \;1\; &\;0\;       \\
0  & 0   &0 & 1      \\
   \hline
\;\half\; & 0 & 0  &0       \\
0 & \;- \half\; & 0  &0      
 \end{array}
\right), 
\quad
M_2 = \left(
\begin{array}{cc|cc}
\;-\nu\; & 0 &  \;0\; &\;0 \;       \\
0 & \;-\eta\;  &0 &0\;       \\
   \hline
0 & 0 & \;-\nu\; &\;iB'_0\;       \\
0 & 0  &\;-iB'_0\; &\;-\eta\;       \\
 \end{array}
\right) . 
\end{equation}

For an approximate growth rate $p$ we use the expansion (\ref{eqbasicM0series}) and solve order by order. At leading order (\ref{eqgenprob0}) we focus on the eigenvalues $p_0 = \pm iB'_0$ of $M_0$, corresponding to large-scale undamped Alfv\'en waves. We will focus on the upper sign without loss of generality, and take
\begin{equation}
p_0 =  iB'_0 , \quad \vv_0 = (1,1,0,0)^T , \quad  \wv_{0} = (1,1,0,0). 
\end{equation}
Here $\wv_0$ is the left eigenvector as usual, with  $\wv_0  (M_0-p_0) = 0 $ and $\wv_0 \vv_0 = 2$. 

Moving to the first order, from (\ref{eqgenprob1a}) we rapidly find 
\begin{equation}
  M_1 \vv_0 = (0,0,\half , -\half)^T,   \quad p_1 = 0 .
\end{equation}
We now need to solve (\ref{eqgenprob1}) for $\vv_1$. To find a solution we clearly need only invert the $2 \times 2$ lower right block of $M_0-p_0$ to calculate
\begin{equation}
\begin{pmatrix} -\nu - iB'_0  & iB'_0 \\ iB'_0 & -\eta - iB'_0 \end{pmatrix}^{-1} 
\begin{pmatrix} - \half \\  \half \end{pmatrix}
= \half \Delta 
\begin{pmatrix}  \eta \\  - \nu \end{pmatrix}
\end{equation}
with the inverse determinant $\Delta$  now defined as 
\begin{equation}
\Delta^{-1} = \eta \nu + iB'_0 (\eta + \nu). 
\end{equation}
Thus 
\begin{equation}
\vv_1 =\half \Delta  (0, 0,  \eta , -  \nu ) ^T.
\end{equation}

With this it is straightforward to calculate $p_2$ from (\ref{eqgenprob2a}) 
\begin{equation}
p_2 =  \tfrac{1}{4} \Delta (\eta - \nu) - \half (\nu + \eta). 
\end{equation}
Taking the real part of $p_2$, and then putting back $B'_0 = k B_0$ and $p = p_0 + k p_1 + k^2 p_2 + \cdots$, gives growth rates 
\begin{equation}
\ReRe p =   \frac{\tfrac{1}{4}   \nu \eta (\eta - \nu )k^2 }{\nu^2 \eta^2 + k^2 B_0^2  (\nu + \eta)^2 } -  \tfrac{1}{2} (\nu + \eta) k^2 + \cdots .
\label{eqstrongvertfieldRep} 
\end{equation}
This is taken up in the main body of the paper as (\ref{eqstrongvertfieldRep0}).




\section{Vertical weak field, with $U_0 \neq 0$, $\ell=0$}

We now continue with our analysis of vertical field instabilities for $\ell=0$ and $U_0\neq 0$. We studied the strong field branch with $B_0 = O(k^{-1})$ in the previous appendix. In the present appendix we will take $B_0 = O(k^2)$: this addresses the branch of vertical field instability as seen in figure \ref{fig-vertical-nu-B0} and allows us to resolve the question of how magnetic field suppresses the purely hydrodynamic instability onset and reduces the critical value of $\nu$ below $\nu_c = 2^{-1/2}$. 
We will see that our results will be correct qualitatively even when $B_0$ is as large as order unity while $k\to0$. Nonetheless it is convenient to refer to this as the `weak field branch' to contrast with the strong field branch in the previous appendix. 

We start with the matrix system (\ref{eqfullvertfieldmatrix}) for instability in the presence of vertical field: $ M \vv = p\vv$. However before expanding $M_0$ in powers of $k$ we first rescale  $B_0 = k^2 B'_0$, and set $\nu = \nu_0 + k^2 \nu_2 + \cdots$, $\eta = \eta_0 + k^2 \eta_2 + \cdots$. Here we are going to develop perturbation theory around the critical point $\nu_c(U_0)$ for the purely hydrodynamic problem, with $\nu_c(0) = 2^{-1/2}$. Expanding in powers of $k$ yields
\begin{equation}
M_0 = \left(
\begin{array}{ccc|ccc}
0 & 0 &  0  &\;\;0\; &0  & 0      \\
\;0 \;& \;-\nu_0 \;& \;-iU_0 \; &\;\;0\; &\;0\;  & \;0\;      \\
0 & -iU_0 & \; -\nu_0\;  &\;\;0\; &0  & 0      \\
  \hline
0 & 0 &  0  &\;\;0\; &0  & 0      \\
0 & 0 &  0  &\;\;0\; &-\eta_0  & -iU_0      \\
0 & 0 &  0  &\;\;0\; &-iU_0  & -\eta_0      \\
\end{array}
\right), \quad
\quad
M_1 = \left(
\begin{array}{ccc|ccc}
\;0 \;& \;1\; &  \;0\;  &\;0\; &\;0\;  & \;0\;      \\
\half  & 0 &  0  &0 &0  & 0      \\
0 & 0 &  0  &0 &0  & 0      \\
  \hline
0 & 0 &  0 \;\; &0 &1  & 0      \\
0 & 0 &  0  &\;-\half  &0  & 0      \\
0 & 0 &  0  &0 &0  & 0      \\
\end{array}
\right), 
\end{equation}
\begin{equation}
M_2 = \left(
\begin{array}{ccc|ccc}
-\nu_0 & 0 &  0  &0 &0  & 0      \\
0 & -\nu_0 -\nu_2 &  0  &0 &0  & 0      \\
0 & 0 &  -\nu_0 -\nu_2\;  &0 &0  & 0      \\
  \hline
0 & 0 &  0  &\;-\eta_0\; &0  & 0      \\
0 & 0 &  0  &0 &-\eta_0 - \eta_2  & 0      \\
0 & 0 &  0  &0 &0  & -\eta_0 - \eta_2      \\
\end{array}
\right), 
\quad
M_3 = \left(
\begin{array}{ccc|ccc}
0 & -1 &  0  &\;iB'_0\; &0  & 0      \\
-\half  & 0 &  0  &0 &\;iB'_0\;  & 0      \\
0 & 0 &  0  &0 &0  & \;iB'_0\;      \\
  \hline
\;iB'_0\; & 0 &  0  &0 &0  & 0      \\
0 & \;iB'_0\; &  0  &0 &0  & 0      \\
0 & 0 & \;iB'_0\;  &0 &0  & 0      \\
\end{array}
\right), 
\end{equation}
\begin{equation}
M_4 = \left(
\begin{array}{ccc|ccc}
-\nu_2 & 0 &  0  &0 &0  & 0      \\
0 & -\nu_2 - \nu_4 &  0  &0 &0  & 0      \\
0 & 0 &  -\nu_2 - \nu_4  &0 &0  & 0      \\
  \hline
0 & 0 &  0  &-\eta_2  & 0 & 0      \\
0 & 0 &  0  &0 &-\eta_2 - \eta_4  & 0      \\
0 & 0 &  0  &0 &0  & -\eta_2 - \eta_4      \\
\end{array}
\right). 
\end{equation}
It will be convenient to let
\begin{equation}
\Delta^{-1} = \nu_0^2 + U_0^2, \quad 
\delta^{-1} = \eta_0^2 + U_0^2. 
\end{equation}

In keeping with the hydrodynamic case we will be expanding the system about a zero eigenvalue $p_0=0$ of $M_0$ with a flow field specified in $G_0$. However we should note that $p_0$ is a twice repeated eigenvalue and we have two left and two right eigenvectors which we distinguish with $\dag$ and $\ddag$: 
\begin{equation}
\vv_{0\dag}^T  =  \wv_{0\dag} = (1,0,0,0,0,0), \quad 
\vv_{0\ddag}^T  = \wv_{0\ddag} = (0,0,0,1,0,0) . 
\end{equation}
In the perturbation theory for a general matrix $M$ we would need to take $\vv_0$ as a general combination of $\vv_{0\dag}$ and $\vv_{0\ddag}$, to be determined further in the expansion. However here, to avoid unnecessary algebra we will jump to the solution we need, and take 
\begin{equation}
p_0 = 0 , \quad \vv_0 \equiv \vv_{0\dag} , 
\end{equation}
for a flow-dominated eigenfunction. We verify that this works as we delve  into the expansion. 

Looking to the first order equation (\ref{eqgenprob1}) we apply $\wv_{0\dag}$ and $\wv_{0\ddag}$ to the left-hand side, which only gives $p_1=0$ and we have 
%
%
\begin{equation}
p_1 = 0 ,\quad M_0 \vv_1 = (0,-\half , 0 , 0,0,0,0)^T. 
\end{equation}
However when we solve for $\vv_1$ we can add not only a multiple of $\vv_0$ to the solution (which  would have no effect in the calculation) but also a multiple of the purely magnetic eigenvector $\vv_{0\ddag}$. Thus we solve for $\vv_1$ in the form  
\begin{equation}
   \vv_1 = (0 ,\half \nu_0\Delta  ,- \half  i U_0\Delta,b,0,0)^T , 
\end{equation}
where $b$ is an unknown constant, to be determined. 

At the next order we will aim to take $p_2 = 0$ so as to push various the effects down the series in powers of $k$: this is achieved if we fix $\Delta = 2$. Thus at second order  we set  
\begin{equation}
p_2 = 0 ,\quad \Delta = 2 , 
\end{equation}
and note that this then fixes 
\begin{equation}
\nu_ 0 \equiv \nu_c (U_0) = ( \half  - U_0^2)^{1/2}. 
\label{eqvwnucfix}
\end{equation}
This is the critical value for onset for the pure hydrodynamic case, with a mean flow but with zero magnetic field. A strong enough mean flow $|U_0|> 2^{-1/2}$ is enough to suppress any instability and so from now on we take $U_0^2<\frac{1}{2}$ so that $\nu_0 $ is real and positive. 

We then need to solve (\ref{eqgenprob2}) which amounts to 
\begin{equation}
 M_0 \vv_2 =(0,0,0,0,\half b  ,0)^T , 
\end{equation}
a suitable solution being 
\begin{equation}
 \vv_2 =  (0,0,0,0, - \half  \eta_0 b\delta , \half   iU_0 b \delta)^T . 
\end{equation}
It turns out that we do not gain any futher information in the ensuing calculation if we incorporate an unknown multiple of $\vv_{0\ddag}$ in our solution for $\vv_2$, at this order, and so we do not.  

At third order, applying $\wv_{0\dag}$ and $\wv_{0\ddag}$ to (\ref{eqgenprob3}) gives 
\begin{equation}
p_3 = 0 ,\quad 
 b  = \eta_0^{-1} (1+ \half  \delta)^{-1} i B'_0, 
 \label{equsefulbeqn}
 \end{equation}
and we then solve 
\begin{equation}
M_0 \vv_3 =( 0, \half + \half (\nu_0 + \nu_2) \nu_0 \Delta , -  \half (\nu_0 + \nu_2) i U_0 \Delta , 0, 0, 0)^T , 
\end{equation}
for $\vv_3$ with  
\begin{equation}
\quad \vv_3 =  (0, - \half \nu_0 \Delta - \half (\nu_0 + \nu_2) ( \nu_0^2 - U_0^2) \Delta^2,  \half i U_0 \Delta +  (\nu_0+\nu_2) i U_0 \nu_0 \Delta^2 ,0,0,0 )^T, 
\end{equation}

Finally applying $\wv_{0\dag}$ and $\wv_{0\ddag}$ to (\ref{eqgenprob4}) gives our desired growth rate 
\begin{equation}
p_4 = - \nu_2 - \nu_0 \Delta+ iB'_0 b - \half (\nu_0 + \nu_2) ( \nu_0^2 - U_0^2) \Delta^2
\end{equation}
We now put $p = p_4 k^4 + \cdots$, and substitute  $b$ from (\ref{equsefulbeqn}), $\Delta=2$, $B_0' = k^{-2} B_0$, $\eta_0 = \nu_0  / P$, also replacing $U_0$ in terms of $\nu_0 = \nu_c$ and 
  $\nu_2 = k^{-2} (\nu - \nu_c) + \cdots$, to find 
%
\begin{equation}
p =  -  \frac{P}{\nu_c(1+\half \delta)} \,  B_0^2  + 4 \nu_c^2 (\nu_c - \nu) k^2 - \nu_c (1+4\nu_c^2) k^4  + \cdots, 
\end{equation}
with 
\begin{equation}
\nu_c = \sqrt{\half - U_0^2}\, ,  \quad
\delta = \frac{P^2}{\nu_c^2( 1 - P^{2}) + \half P^2 }\, . 
\end{equation}
This is the general formula for the growth rate $p(k, \nu, B_0, P, U_0)$ including a mean flow $U_0$ with $U_0^2< \half$ (so that $\nu_c$ is defined as a positive number). 

In the case of zero mean flow, $U_0=0$, we have $\nu_c = 2^{-1/2}$, $\delta = 2P^2$,  and this simplifies to
\begin{equation}
p =  -  \frac{\sqrt{2}\, P}{1+P^2} \,   B_0^2  +2 \biggl(\frac{1}{\sqrt{2}} - \nu\biggr) k^2 - \frac{3}{\sqrt{2}}\,  k^4  + \cdots. 
\label{eqappCp}
\end{equation}
We continue the discussion in the main body of the paper; see equations (\ref{eqappCpmax}, \ref{eqappCpthreshold}). Note that going to fourth order here suggests that the modes $G_{\pm2}$ and $H_{\pm2}$ might also be needed to give a correct evaluation of $p_4$; this needed to be checked and was --- we found that the couplings are too weak, in terms of the powers of $k$ involved, to give a contribution to the growth rate to the order taken above. 




\section{Horizontal field, with $U_0=0$, $\ell \neq0$}

Our final calculation brings in the Bloch wavenumber $\ell$. This can be done in the case of vertical field \citep{mephd2023}, but there increasing $\ell$ from zero seems to have only the effect of suppressing the $\ell=0$ instability, so we do not consider this further. We instead study horizontal field, where having $\ell\neq0$ can enhance instability. We take $U_0=0$ to keep the problem manageable. We omit straightforward but messy details and in fact will only go up to first order in perturbation theory.

Our starting point is the equations (\ref{eqGnhoriz}, \ref{eqHnhoriz}) with $n$ replaced by $n+\ell$, and we consider only the modes $G_0$, $G_{\pm1}$, $H_0$, $H_{\pm1}$. We set, as in the original horizontal field problem (appendix A), 
\begin{equation}
G_0 = k^{2}G'_0 , \quad
\Bt_0 = B_0 / \eta. 
\end{equation}
Once we have $\ell \neq0$ in the problem, we have to ask how $\ell$ scales as $k\to0$. It turns out that the appropriate scaling to gain useful results is 
\begin{equation}
\ell = k \ell'  = O(k) ,
\end{equation}
so we hold $\ell'$ and $G'_0$ constant while $k \to 0$. We now follow the usual procedure of making these substitutions, expressing the equations in terms of $G_0$, $G_{\pm}$, $H_0$, $H_{\pm}$ and writing the system as $M\vv = p \vv$ with 
\begin{equation}
\vv =  \begin{pmatrix} G'_{0} \\ G_{-} \\ H_{+} \\ H_{0} \\ H_{-} \\G_{+}\\ \end{pmatrix}, 
\end{equation}
and $M = M_0 + k M_1 + \cdots$ with 
\begin{equation}
\Mv_0 = 
\left(
\begin{array}{ccc|ccc}
\;0\; &\; 0\; & \;0 \;& \;\;0\;  &  \;2 \ell'  i\Bt_0\; & \;2 \ell'\;  \\
0 & \;-\nu\; & \; i B_0\; & 0 &  0 &  0 \\
0 &  \;iB_0\; & \; -\eta\; & 0 & 0 & 0 \\
  \hline
 0& 0  & 0 & 0  &  0 & 0  \\
0 & 0 & 0 & 0  & \;- \eta\; & \;iB_0 \;  \\
0 &  0 &   0  &0  & \;i B_0 \;& \;-\nu\; \\
\end{array}
\right), 
\end{equation}
and
\begin{equation}
\Mv_1 = \left(
\begin{array}{ccc|ccc}
0&\; 1 - 3 \ell'^2\; & \; i\Bt_0 (1+\ell'^2)\;  & \;iB_0 \ell' (1+\ell'^2)\;  & 0 & 0 \\
\;\half  (1 + \ell'^2)^{-1} \; & 0 &  0 & 0 & \; iB_0 3 \ell'\; & \; - 2  \ell' \nu \;\\
\;\half i\Bt_0  (1 + \ell'^2)^{-1}\; & 0&  0& 0 & \;- 2 \ell' \eta\; & \;- i B_0 \ell'\; \\[2pt]
\hline
 iB_0  {\ell'}(1+\ell'^2)^{-1} & 0  & 0 & 0 &  1 & i\Bt_0  \\
0 & - i B_0 \ell' & - 2 \ell' \eta & -\half   & 0 & 0  \\
0 &  - 2  \ell' \nu &  iB_0 3 \ell'  & - \half {i\Bt_0}  & 0 &0 \\
\end{array}
\right) . 
\end{equation}
It is convenient to set 
\begin{equation}
\Delta^{-1} = \nu \eta + B_0^2 . 
\end{equation}

We are now ready to calculate $p$. The matrix $M_0$ has now lost the attractive block structure present in the earlier expansions as a consequence of the scaling of $\ell$. Nonetheless $M_0$ has a double zero eigenvalue $p_0=0$ with right eigenvectors 
\begin{equation}
  \vv_{\dag 0} = (1,0,0,0,0,0)^T , \quad
    \vv_{\ddag 0} = (0,0,0,1,0,0)^T , 
\end{equation}
and left eigenvectors 
\begin{equation} 
 \wv_{0\dag} = (1,0,0,0,2\ell' \Delta \eta^{-1} iB_0 (\nu+\eta), 2\ell' \Delta \eta^{-1} ( \eta^2- B_0^2) ), \quad \wv_{0\ddag} = (0,0,0,1,0,0) . 
\end{equation}
In the previous case of a double-zero eigenvalue (in appendix C) we anticipated the structure of $\vv_0$ (as dominated by the hydrodynamic field $G_0$). Here we cannot do so and so we set
\begin{equation}
\vv_0 = b \vv_{0\dag} + c \vv_{0\ddag}, 
\end{equation}
for some constants $b$ and $c$. 

Now, looking at the first order equation (\ref{eqgenprob1}), namely $p_1 \vv_0  = (M_0 - p_0) \vv_1 + M_1 \vv_0$ with $p_0=0$, we can apply either of the two vectors 
 $\wv_{0\dag}$ and $\wv_{\ddag}$ on the left, to gain two equations,
\begin{align}
p_1b & = \wv_{0\dag} M_1 \vv_0 = i B_0 \ell' [ 1+\ell'^2 + \Delta \eta^{-2} (B_0^2 - \nu \eta - 2 \eta^2) ] c, \\
p_1 c  & = \wv_{0\ddag} M_1 \vv_0 = iB_0 \ell' (1+\ell'^2)^{-1} b .
\end{align}
Together, these yield
\begin{equation}
p_1^2 = - B_0^2 \ell'^2 [ 1 +  (1+\ell'^2)^{-1} \Delta \eta^{-2} (B_0^2 - \nu \eta - 2 \eta^2) ],
\end{equation}
and so, putting back $\ell' = \ell / k$ and $\Delta$ we find the growth rate as 
\begin{equation}
p =   \pm B_0 \ell \left[ \frac{k^2}{\ell^2+k^2} \, \frac{ \nu \eta + 2 \eta^2 - B_0^2}{\eta^2 (\nu\eta + B_0^2)} -  1 \right]^{1/2} + \cdots . 
\label{eqhorizellp}
\end{equation}
We have gained this equation by only going to the first order matrix $M_1$, but it reveals an instability that crucially relies on having a non-zero Bloch wavenumber, $\ell\neq0$. We continue our discussion in the main body of the paper, from (\ref{eqhorizellpmain}).



\section*{Acknowledgements}
\noindent
ADG acknowledges support from the EPSRC Research Grant EP/T023139/1 and AH acknowledges support from the STFC Research Grant ST/V000659/1.  AMA is grateful for a PhD studentship awarded by the government of Saudi Arabia. For the purpose of open access, the author has applied a Creative Commons Attribution (CC BY) licence to any Author Accepted Manuscript version arising. 

\section*{Data access statement}
\noindent
No data were created or analysed in this study.


\bibliographystyle{gGAF2e}
\bibliography{ref}

\begin{thebibliography}{30}
\providecommand{\natexlab}[1]{#1}

\bibitem[\protect\citeauthoryear{Algatheem}{2023}]{mephd2023}
Algatheem, A.M., Jets and instabilities in forced MHD flows, in preparation.
  Ph.D. Thesis, University of Exeter Unpublished thesis, 2023.

\bibitem[\protect\citeauthoryear{Balmforth and
  Young}{2002}]{balmforth2002stratified}
Balmforth, N.J. and Young, Y.N., Stratified Kolmogorov flow. {\itshape Journal
  of Fluid Mechanics}, 2002, \textbf{450}, 131--167.

\bibitem[\protect\citeauthoryear{Boffetta
  {\itshape{et~al.}}}{2000}]{boffetta2000large}
Boffetta, G., Celani, A. and Prandi, R., Large scale instabilities in
  two-dimensional magnetohydrodynamics. {\itshape Physical Review E}, 2000,
  \textbf{61}, 4329.

\bibitem[\protect\citeauthoryear{Bouchet
  {\itshape{et~al.}}}{2013}]{bouchet2013kinetic}
Bouchet, F., Nardini, C. and Tangarife, T., Kinetic theory of jet dynamics in
  the stochastic barotropic and 2D Navier-Stokes equations. {\itshape Journal
  of Statistical Physics}, 2013, \textbf{153}, 572--625.

\bibitem[\protect\citeauthoryear{Chechkin}{1999}]{chechkin1999negative}
Chechkin, A., Negative magnetic viscosity in two dimensions. {\itshape Journal
  of Experimental and Theoretical Physics}, 1999, \textbf{89}, 677--688.

\bibitem[\protect\citeauthoryear{Constantinou and
  Parker}{2018}]{constantinou2018magnetic}
Constantinou, N.C. and Parker, J.B., Magnetic suppression of zonal flows on a
  beta plane. {\itshape The Astrophysical Journal}, 2018, \textbf{863}, 46.

\bibitem[\protect\citeauthoryear{Dubrulle and Frisch}{1991}]{dubrulle1991eddy}
Dubrulle, B. and Frisch, U., Eddy viscosity of parity-invariant flow. {\itshape
  Physical Review A}, 1991, \textbf{43}, 5355.

\bibitem[\protect\citeauthoryear{Durston and
  Gilbert}{2016}]{durston2016transport}
Durston, S. and Gilbert, A.D., Transport and instability in driven
  two-dimensional magnetohydrodynamic flows. {\itshape Journal of Fluid
  Mechanics}, 2016, \textbf{799}, 541--578.

\bibitem[\protect\citeauthoryear{Farrell and
  Ioannou}{2008}]{farrell2008formation}
Farrell, B.F. and Ioannou, P.J., Formation of jets by baroclinic turbulence.
  {\itshape Journal of the Atmospheric Sciences}, 2008, \textbf{65},
  3353--3375.

\bibitem[\protect\citeauthoryear{Fraser
  {\itshape{et~al.}}}{2022}]{fraser2022non}
Fraser, A., Cresswell, I. and Garaud, P., Non-ideal instabilities in sinusoidal
  shear flows with a streamwise magnetic field. {\itshape Journal of Fluid
  Mechanics}, 2022, \textbf{949}, A43.

\bibitem[\protect\citeauthoryear{Frisch
  {\itshape{et~al.}}}{1996}]{frisch1996large}
Frisch, U., Legras, B. and Villone, B., Large-scale Kolmogorov flow on the
  beta-plane and resonant wave interactions. {\itshape Physica D: Nonlinear
  Phenomena}, 1996, \textbf{94}, 36--56.

\bibitem[\protect\citeauthoryear{Galperin
  {\itshape{et~al.}}}{2006}]{galperin2006anisotropic}
Galperin, B., Sukoriansky, S., Dikovskaya, N., Read, P., Yamazaki, Y. and
  Wordsworth, R., Anisotropic turbulence and zonal jets in rotating flows with
  a $\beta$-effect. {\itshape Nonlinear Processes in Geophysics}, 2006,
  \textbf{13}, 83--98.

\bibitem[\protect\citeauthoryear{Hughes
  {\itshape{et~al.}}}{2007}]{hughes2007solar}
Hughes, D.W., Rosner, R. and Weiss, N.O., {\itshape The solar tachocline},
  2007 (Cambridge University Press).

\bibitem[\protect\citeauthoryear{Kim}{2007}]{kim2007role}
Kim, E.j., Role of magnetic shear in flow shear suppression. {\itshape Physics
  of Plasmas}, 2007, \textbf{14}, 084504.

\bibitem[\protect\citeauthoryear{Leprovost and
  Kim}{2009}]{leprovost2009turbulent}
Leprovost, N. and Kim, E.j., Turbulent transport and dynamo in sheared
  magnetohydrodynamics turbulence with a nonuniform magnetic field. {\itshape
  Physical Review E}, 2009, \textbf{80}, 026302.

\bibitem[\protect\citeauthoryear{Lucas and
  Kerswell}{2014}]{lucas2014spatiotemporal}
Lucas, D. and Kerswell, R., Spatiotemporal dynamics in two-dimensional
  Kolmogorov flow over large domains. {\itshape Journal of Fluid Mechanics},
  2014, \textbf{750}, 518--554.

\bibitem[\protect\citeauthoryear{Lucas and Kerswell}{2015}]{lucas2015recurrent}
Lucas, D. and Kerswell, R.R., Recurrent flow analysis in spatiotemporally
  chaotic 2-dimensional Kolmogorov flow. {\itshape Physics of Fluids}, 2015,
  \textbf{27}, 045106.

\bibitem[\protect\citeauthoryear{Manfroi and
  Young}{2002}]{Manfroi2002stability}
Manfroi, A. and Young, W., Stability of $\beta$-plane Kolmogorov flow.
  {\itshape Physica D: Nonlinear Phenomena}, 2002, \textbf{162}, 208--232.

\bibitem[\protect\citeauthoryear{Meshalkin and
  Sinai}{1961}]{meshalkin1961investigation}
Meshalkin, L. and Sinai, I.G., Investigation of the stability of a stationary
  solution of a system of equations for the plane movement of an incompressible
  viscous liquid. {\itshape Journal of Applied Mathematics and Mechanics},
  1961, \textbf{25}, 1700--1705.

\bibitem[\protect\citeauthoryear{Nepomniashchii}{1976}]{nepomniashchii1976stability}
Nepomniashchii, A., On stability of secondary flows of a viscous fluid in
  unbounded space: PMM vol. 40, no. 5, 1976, pp. 886--891. {\itshape Journal of
  Applied Mathematics and Mechanics}, 1976, \textbf{40}, 836--841.

\bibitem[\protect\citeauthoryear{Parker and
  Constantinou}{2019}]{parker2019magnetic}
Parker, J.B. and Constantinou, N.C., Magnetic eddy viscosity of mean shear
  flows in two-dimensional magnetohydrodynamics. {\itshape Physical Review
  Fluids}, 2019, \textbf{4}, 083701.

\bibitem[\protect\citeauthoryear{Parker and
  Krommes}{2014}]{parker2014generation}
Parker, J.B. and Krommes, J.A., Generation of zonal flows through symmetry
  breaking of statistical homogeneity. {\itshape New Journal of Physics}, 2014,
  \textbf{16}, 035006.

\bibitem[\protect\citeauthoryear{Read
  {\itshape{et~al.}}}{2007}]{read2007dynamics}
Read, P.L., Yamazaki, Y.H., Lewis, S.R., Williams, P.D., Wordsworth, R.,
  Miki-Yamazaki, K., Sommeria, J. and Didelle, H., Dynamics of convectively
  driven banded jets in the laboratory. {\itshape Journal of the Atmospheric
  Sciences}, 2007, \textbf{64}, 4031--4052.

\bibitem[\protect\citeauthoryear{Scott and
  Dritschel}{2012}]{scott2012structure}
Scott, R.K. and Dritschel, D.G., The structure of zonal jets in geostrophic
  turbulence. {\itshape Journal of Fluid Mechanics}, 2012, \textbf{711},
  576--598.

\bibitem[\protect\citeauthoryear{She}{1987}]{she1987metastability}
She, Z.S., Metastability and vortex pairing in the Kolmogorov flow. {\itshape
  Physics Letters A}, 1987, \textbf{124}, 161--164.

\bibitem[\protect\citeauthoryear{Sivashinsky}{1985}]{sivashinsky1985weak}
Sivashinsky, G.I., Weak turbulence in periodic flows. {\itshape Physica D:
  Nonlinear Phenomena}, 1985, \textbf{17}, 243--255.

\bibitem[\protect\citeauthoryear{Srinivasan and
  Young}{2012}]{srinivasan2012zonostrophic}
Srinivasan, K. and Young, W., Zonostrophic instability. {\itshape Journal of
  the Atmospheric Sciences}, 2012, \textbf{69}, 1633--1656.

\bibitem[\protect\citeauthoryear{Tobias
  {\itshape{et~al.}}}{2011}]{tobias2011astrophysical}
Tobias, S., Dagon, K. and Marston, J., Astrophysical fluid dynamics via direct
  statistical simulation. {\itshape The Astrophysical Journal}, 2011,
  \textbf{727}, 127.

\bibitem[\protect\citeauthoryear{Tobias
  {\itshape{et~al.}}}{2007}]{tobias2007beta}
Tobias, S.M., Diamond, P.H. and Hughes, D.W., $\beta$-plane magnetohydrodynamic
  turbulence in the solar tachocline. {\itshape The Astrophysical Journal},
  2007, \textbf{667}, L113.

\bibitem[\protect\citeauthoryear{Vallis and
  Maltrud}{1993}]{vallis1993generation}
Vallis, G.K. and Maltrud, M.E., Generation of mean flows and jets on a beta
  plane and over topography. {\itshape Journal of Physical Oceanography}, 1993,
  \textbf{23}, 1346--1362.

\end{thebibliography}

\end{document}